\documentclass[12pt]{iopart}

\usepackage[dvipdfmx]{graphicx}
\usepackage{color}

\usepackage{comment}

\begin{document}

\title[Site-selective Co substitution in M-type ferrites]{Site-selective cobalt substitution 
in La--Co co-substituted magnetoplumbite-type ferrites:
 $^{59}$Co-NMR and DFT calculation study
}

\author{Hiroyuki Nakamura$^{1}$, Hiroto Ohta$^{2}$, Ryuya Kobayashi$^1$, Takeshi Waki$^{1}$, Yoshikazu Tabata$^{1}$, Hidekazu Ikeno$^3$ and Christian M\'eny$^4$}
\address{$^1$ Department of Materials Science and Engineering, Kyoto University, Kyoto 606-8501, Japan}
\address{$^2$ Department of Molecular Chemistry and Biochemistry, Faculty of Science and Engineering, Doshisha University, Kyotanabe 610-0321, Japan}
\address{$^3$ Department of Materials Science, Graduate School of Engineering, Osaka Metropolitan University, Sakai, Osaka, 599-8570, Japan}
\address{$^4$ Institut de Physique et Chimie des Mat\'eriaux de Strasbourg (IPCMS), UMR7504, CNRS-Universit\'e de Strasbourg, 23, rue du Loess, F-67034 Strasbourg Cedex 02, France}
\ead{nakamura.hiroyuki.2w@kyoto-u.ac.jp}

\begin{abstract}
 The La--Co co-substituted magnetoplumbite-type (M-type) ferrites $A$Fe$_{12}$O$_{19}$ ($A$ = Ca, Sr and Ba, ion sizes Ca$^{2+}$ $<$ Sr$^{2+}$ $<$ Ba$^{2+}$) with Co compositions around 0.2 have been subjected to $^{59}$Co-NMR.
The results show that Co occupies the 4f$_1$, 2a and 12k sites, and that the smaller the $A$ ion, the more Co tends to occupy the 4f$_1$ minority spin site, which is effective in enhancing both uniaxial anisotropy and magnetisation.
First-principles total energy calculations based on density functional theory (DFT) of undoped $A$Fe$_{12}$O$_{19}$ and a supercell ($2 \times 2 \times 1$ of the unit cell) in which 1/96 of Fe$^{3+}$ is replaced by Co$^{2+}$ were performed to predict the stable structure and Co occupancy sites.
The results show that regardless of $A$, Co is most stable when it occupies the 4f$_1$ site, followed by the 2a and 12k sites with energy differences on the order of 100 meV, and Co practically does not occupy the 2b and 4f$_2$ sites.
As the $A$ ion becomes smaller, the energy difference when Co occupies each Fe site tends to increase, and the Co  occupancy of the 4f$_1$ site also increases.
The site selectivity of Co can be roughly explained as a result of the difference in uniaxial strain along the $c$-axis associated with the difference in $A$.
However, the influence of the $A$ ion differs between the R and S blocks and the local strain also has a secondary effect on the Co distribution.
Based on these results, the guidelines for improving the performance (anisotropy and magnetisation) of La--Co co-substituted M-type ferrite magnets with a limited amount of Co can be summarised as follows: It is effective to select as small $A$ ions as possible and to post-anneal at low temperature or cool slowly to concentrate Co at the 4f$_1$ site in tetrahedral coordination.
\end{abstract}


\noindent{\it Keywords}: ferrite magnet, anisotropy, NMR, DFT calculation, magnetoplumbite-type ferrite
%
%
%
%

\begin{figure}[t]
\begin{center}
\includegraphics[width=0.45\textwidth]{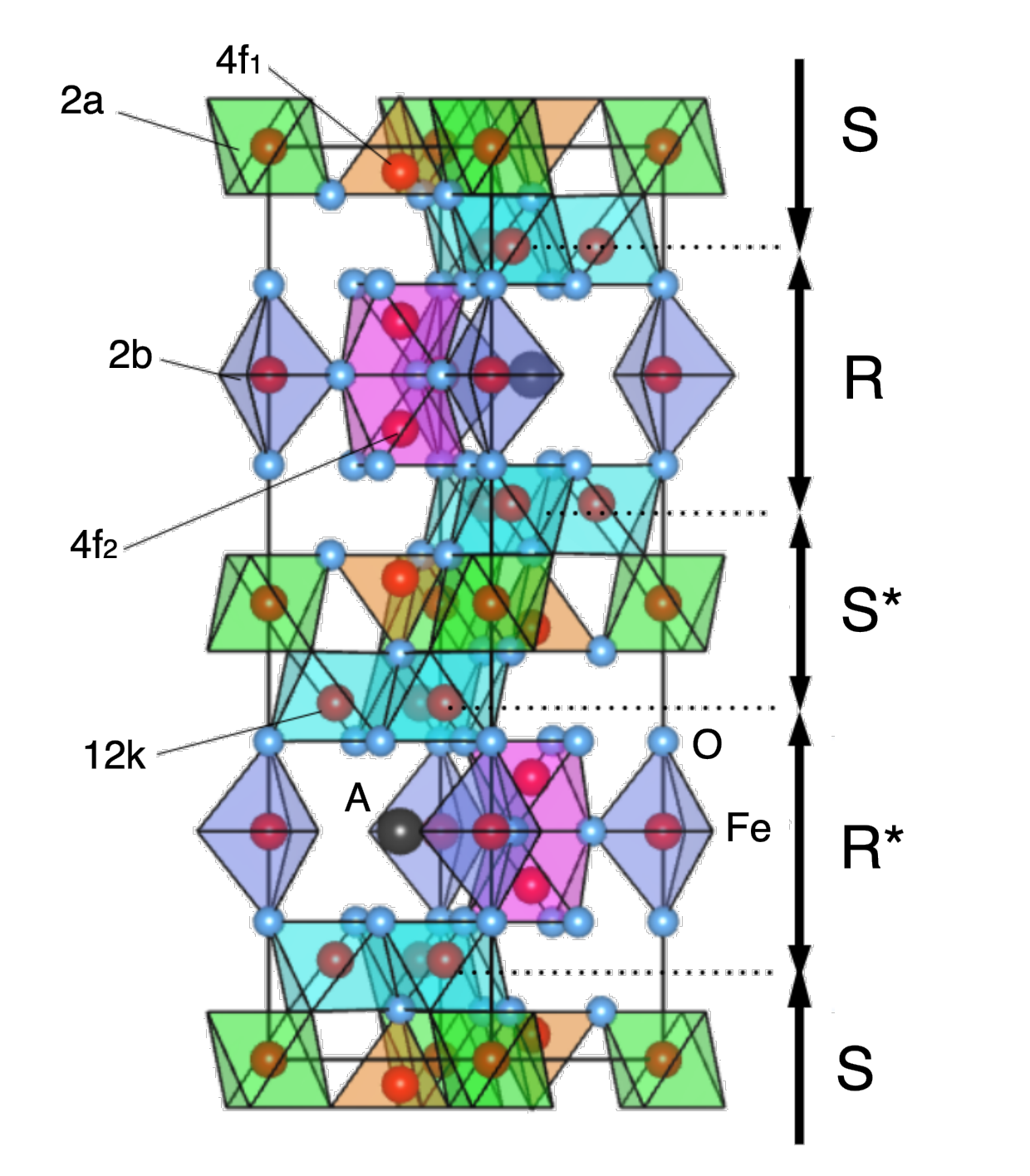}
\caption{\label{fig_crystal}
Crystal structure of the magnetoplumbite-type (M-type) $A$Fe$_{12}$O$_{19}$ drawn by VESTA \cite{Momma2011}.
The space group is $P6_3/mmc$.
The unit cell is equal to two formula units.
The $A$ ion (black circle) occupies the 2d site surrounded by 12 oxygen ions.
Iron (red circles) has five crystallographically inequivalent sites (2a, 2b, 4f$_1$, 4f$_2$ and 12k) and its oxygen coordination polyhedra are represented by different colours.
The 2a, 4f$_2$ and 12k sites are 6-coordinated (octahedral), the 4f$_1$ site is 4-coordinated (tetrahedral) and the 2b site is 5-coordinated (bipyramidal).
The 2a, 2b, 4f$_1$ and 4f$_2$ sites have trigonal rotational symmetry around the $c$-axis, but the 12k site is not trigonally symmetric, with the local principal axis tilted from the $c$-axis resulting in three types of 12k-Fe.
Oxygen (blue circles) has five crystallographically inequivalent sites (4e, 4f, 6h, 12k$_1$, 12k$_2$).
All Fe is trivalent and is expected to have a spin moment of 5 $\mu_\mathrm{B}$/Fe.
$A$Fe$_{12}$O$_{19}$ is ferrimagnetic; the Fe at the 2a, 2b and 12k sites has the majority spin and the Fe at the 4f$_1$ and 4f$_2$ sites has the minority spin.
S and R denote \textit{blocks} and the M-type structure has SRS$^*$R$^*$ stacking, where * denotes a 180$^\circ$ rotation around the $c$-axis.
}
\end{center}
\end{figure}

\section{Introduction}
\label{sec_intro}

Ferrite magnets are used in a wide range of equipment because they are extremely cost effective compared with rare earth magnets \cite{Pullar2012}.
Higher performance and lower cost of ferrite magnets are the demands of society.
Many attempts have been made to improve the performance of ferrite magnets \cite{Julian2021}. 
Many of these are elemental substitutions, and the substitution of Co for Fe is a successful example \cite{Iida1999}.
However, much of the development is trial and error and not necessarily based on clear guiding principles.
Against this background, the aim of this study was to propose guidelines beyond trial and error for the development of the next generation of high performance magnets and the effective use of resources.
To improve the performance of magnetoplumbite-type (M-type) ferrites, improvements in magnetic anisotropy (coercivity), magnetisation (remanence), or both, are required.
This paper focuses on methods to improve magnetic anisotropy without degrading or even improving magnetisation.

The base material of ferrite magnets is the hexagonal M-type ferrite $A$Fe$_{12}$O$_{19}$ ($A$ = Ba, Sr, Pb...) (figure \ref{fig_crystal}).
The space group of the M-type structure is $P6_3/mmc$, and the unit cell contains two formula units.
All Fe atoms are located in the gap between O and $A$.
There are five crystallographically inequivalent Fe sites in the Wyckoff symbols: 2a (octahedral coordination), 2b (bipyramidal), 4f$_1$ (tetrahedral), 4f$_2$ (octahedral) and 12k (octahedral).
The M-type structure can be divided into the S block (Fe$_6$O$_8$)$^{2+}$ including 2a and 4f$_1$ sites and the R block ($A$Fe$_6$O$_{11}$)$^{2-}$ including 2b and 4f$_2$ sites, with the 12k site in the plane perpendicular to the $c$-axis at the boundary between the S and R blocks.
The stacking is described as SRS$^*$R$^*$, where * represents a 180$^\circ$ rotation around the $c$-axis.
All Fe ions in $A$Fe$_{12}$O$_{19}$ are trivalent (3d$^5$) and have a spin magnetic moment of 5 $\mu_\mathrm{B}$ ($S = \frac{5}{2}$).
The M-type ferrites are ferrimagnets in which the magnetic moment of Fe$^{3+}$ is ordered by superexchange interactions through O$^{2-}$.
The 2a, 2b and 12k sites are the majority spin sites, and the 4f$_1$ and 4f$_2$ sites are the minority spin sites, resulting in a net magnetic moment of $ 20~\mu_\mathrm{B}$ per formula unit. 
Isolated Fe$^{3+}$ with no orbital component in the magnetic moment do not normally cause magnetic anisotropy. 
The presence of the asymmetric 2b site, which exhibits strong single-ion anisotropy \cite{Fuchikami1965} via interaction with the coordinating oxygen ions, appears to be essential for \textit{uniaxial} anisotropy in M-type ferrites. 
However, contributions other than the 2b site are also important for a quantitative explanation of the anisotropy, as has been better understood by recent theoretical work \cite{Inoue2019, Inoue2020, Bhandari2021}.
 
 Various element substitutions have been attempted to improve the performance of M-type ferrite magnets.
 Among these, La--Co co-substituted M-type SrFe$_{12}$O$_{19}$, where some of the Fe$^{3+}$ (approximately 2--3\%) in SrFe$_{12}$O$_{19}$ is replaced by Co$^{2+}$ and some of the Sr$^{2+}$ by La$^{3+}$ as charge compensation, has been reported to improve the coercivity by approximately 30\% and the residual magnetisation by a few percent compared to the undoped material \cite{Iida1999}.
  The residual magnetisation is slightly improved because Co$^{2+}$, which has a smaller magnetic moment than Fe$^{3+}$, mainly occupies the minority spin sites, thereby increasing the net magnetisation.
  It is easy to imagine that the increase in coercivity is due to the enhanced uniaxial magnetic anisotropy caused by the residual (or induced) orbital magnetic moment of Co$^{2+}$.
 However, the mechanism of enhancement has long been unclear due to the ambiguity of the Co occupancy sites.
 However, recent studies have shown that Co$^{2+}$ mainly occupies the 4f$_1$ site and partially occupies the 2a and 12k sites.
 The fact that Co occupies three sites at 2a, 4f$_1$ and 12k was first proposed by a combined Rietveld analysis of neutron diffraction, extended X-ray absorption fine structure (EXAFS) and X-ray magnetic circular dichroism (XMCD) \cite{Kobayashi2011}.
 This result was supported by the atom location determined using the channeling-enhanced microanalysis (ALCHEMI) method \cite{Ohtsuka2016}.
  Recent NMR \cite{Sakai2018} and M\"ossbauer experiments \cite{Oura2018, Nagasawa2020} are consistent with the site occupancy of Co.
 
La--Co co-substituted M-type ferrite magnets with $A$ ions other than Sr have also been developed.
It is known that the case of $A$ = Ca has a higher coercivity than $A$ = Sr, even with the same amount of Co substitution \cite{Kobayashi2008}.
In contrast, the coercivity for $A$ = Ba is lower than for $A$ = Sr \cite{Li2013}.
This suggests that even with the same amount of Co, the Co distribution is not the same and depends on the $A$ ion.
The Co distribution in La--Co co-substituted CaFe$_{12}$O$_{19}$ has also been studied by a combined analysis of Rietveld refinement of neutron diffraction, EXAFS and XMCD \cite{Kobayashi2016, Kobayashi2016T, Kobayashi2017}.
The occupancy sites of Co are the same as in SrFe$_{12}$O$_{19}$, but the 4f$_1$-site occupancy of Co has been reported to be greater in the Ca--La ferrite.

It should also be noted that physical properties such as coercivity can vary greatly depending on the synthesis  process and heat treatment conditions, even for materials of the same composition.
Such cases arise when dopants occupy several non-equivalent sites and their stability is antagonistic at relatively low energy scales.
This is because the cation distribution at room temperature is not necessarily the ideal thermal equilibrium state at room temperature, but is more or less a frozen state of the high temperature distribution.
In other words, the final physical properties depend on the temperature at which the cation distribution is frozen, that is, the heat treatment applied to the sample.
Spinel compounds have long been the focus of studies of cation distributions at finite temperatures \cite{Ma2019}.
Some simple models commonly discuss the temperature dependence of the cation distribution from the probability of occupancy via the Gibbs free energy \cite{Ma2019}.
In recent years, many examples have been reported where density functional theory (DFT) has been used to calculate the energy change of a state in which a particular site is replaced by a particular element, which is then used to calculate the probability of occupancy at a finite temperature \cite{Liu2014, Wu2016, Dixit2017}.
The application of this method to M-type ferrites is growing rapidly \cite{Wu2016, Dixit2017}, and similar discussions can be made in this study.
 
$^{59}$Co ferromagnetic nuclear resonance experiments are a convenient way to obtain information on the site-to-site distribution of Co in ferromagnets and microscopic magnetism, provided that the occupancy sites can be assigned to the resonance signals.
We performed $^{59}$Co-NMR experiments on samples of La--Co co-substituted SrFe$_{12}$O$_{19}$ with widely varying Co compositions (from 0.15 to 0.93 Co) to investigate the Co distribution over multiple Fe sites and the role of Co occupying each site  \cite{Nakamura2019}. 
The results showed that Co was commonly distributed over three Fe sites (4f$_1$, 2a and 12k), with the 4f$_1$ site being the most preferentially occupied.
We evaluated the Co content dependence of the local anisotropy at each site by measuring the signal enhancement factor, and found that the macroscopic anisotropy was positively correlated with the local anisotropy of Co at the 4f$_1$ site and inversely correlated with those at the 2a and 12k sites: see the second paragraph of section 3.2 of \cite{Nakamura2019} for details, and \cite{Meny2021, Panissod2020} for this methodology.
Based on these results, we proposed for the first time that only Co at the 4f$_1$ site contributes to the enhancement of uniaxial anisotropy.
Although no orbital degeneracy of Co$^{2+}$ at the tetrahedrally coordinated site would be expected if the coordination polyhedron were a regular tetrahedron, we concluded that the Co at the 4f$_1$ site, which is trigonally symmetric, is the main cause of the enhancement of uniaxial anisotropy.
From these results, we proposed that the magnetic anisotropy and saturation magnetisation of ferrite magnets can be improved by concentrating Co at the 4f$_1$ site, and the production cost can also be reduced by the effective use of Co.
However, the actual method for concentrating Co at the 4f$_1$ site remains debatable.
In La--Co co-substituted SrFe$_{12}$O$_{19}$, the 4f$_1$-site occupation ratio of Co was unexpectedly increased with increasing La--Co substitution.
It was suggested that the 4f$_1$-site selectivity of Co could be enhanced by a decrease in $c/a$, that is, compressive strain along the $c$-axis, associated with the substitution of La$^{3+}$, which is smaller in size than Sr$^{2+}$. 
However, as a series of samples with very different Co concentrations were used for the measurements, we could not rule out the possibility that the Co distribution changes due to changes in the Co concentration itself and could not clearly conclude that uniaxial strain affects Co site selectivity.

In this study, we focused on the change in Co distribution in La--Co co-substituted $A$Fe$_{12}$O$_{19}$ due to the difference in $A$-ion size.
In other words, similar to the previous study \cite{Nakamura2019}, $^{59}$Co-NMR experiments were performed on La--Co co-substituted $A$Fe$_{12}$O$_{19}$ ($A$ = Ca, Sr and Ba) with a common Co composition to compare the Co site selectivity and investigate the effect of $A$-ion size.
We also evaluated the stability of Co occupying each Fe site and the crystal structure parameters of structure-optimised $A$Fe$_{12}$O$_{19}$ and Co substituted $A$Fe$_{12}$O$_{19}$ by first-principles total energy calculations based on DFT and discussed the agreement with experiments, as well as the factors controlling the Co site selectivity.
As mentioned above, to further improve the coercivity and magnetisation of La--Co co-substituted M-type ferrite magnets and to make effective use of Co, it is necessary to concentrate Co$^{2+}$ more efficiently at the 4f$_1$ site. 
Establishing a simple guiding principle for this will accelerate the development of next-generation high performance ferrite magnets.
In this study, we show that the selection of $A$ ions that reduce the $c$-axis length is effective for this purpose. 
We also discuss the secondary effect of the local strain on Co site selectivity.
From the results of DFT calculations, we show that the Co site-to-site distribution is temperature sensitive in the temperature range where the M phase is stable, and that post-annealing at low temperature and/or slow cooling is effective in concentrating Co at the 4f$_1$ site.

\section{Methods}

\subsection{Our strategy}
\label{sec_strategy}

Since the development of La--Co co-substituted high-performance ferrite magnets \cite{Iida1999}, the mechanism of  coercivity (uniaxial magnetic anisotropy) enhancement has been extensively studied, but the debate has long remained confused.
The main reason for this confusion was that the Co occupancy sites were experimentally ambiguous \cite{Morel2005}.
The amount of Co substituted for Fe is at most a few percent of Fe, and Co is distributed over several sites in five crystallographically inequivalent Fe sites.
In other words, the identification of Co occupancy sites in La--Co co-substituted M-type ferrite is an ultra-trace element analysis in a multi-site compound and requires extremely high experimental precision.
Despite these difficulties, early studies paid insufficient attention to sample homogeneity, compositional reliability, experimental precision and analytical validity.

\subsubsection{Sample}
\label{sec_samplestr}

In many previous studies of M-type ferrites, the reagents were mixed to a certain composition and then calcined, and the initial composition was taken as the nominal composition.
Compositional analysis after synthesis was rarely performed.
This is because the main objective of these studies was to achieve high performance.
In such samples, the actual composition rarely matches the initial composition due to the formation of secondary phases, precipitation of different phases at grain boundaries and segregation of dopants on grain surfaces. Permanent magnets are composite materials that skilfully exploit the different functions of crystal grains and grain boundaries.
The direct use of such materials is inappropriate for evaluating the intrinsic properties of the matrix phase.
In fact, $^{59}$Co-NMR signals of various secondary phases have been observed in NMR experiments on La--Co co-substituted M-type ferrite magnets, making identification of the essential signal difficult.
It is also known that Co segregates on the grain surface depending on the heat treatment conditions and plays a different role from that in the grain \cite{Yamamoto2019}. 
More importantly, the valence of transition metals in M-type ferrite is generally unstable.
La--Co co-substituted M-type ferrites were originally designed to have equal amounts of La and Co, with all Co being divalent, but in reality the divalent (3d$^6$) and trivalent (3d$^5$) states of Fe are energetically antagonistic and generally coexist \cite{Shimoda2016, Tenaud2004}.
The abundance of Fe$^{2+}$ and Fe$^{3+}$ is dependent on the oxygen potential during synthesis and requires a sufficiently high oxygen potential to maintain all Fe in the trivalent state \cite{Waki2018, Waki2023a}.
In other words, if a sample is synthesised at a low oxygen potential, the formation of Fe$^{2+}$ is inevitable.
This results in a discrepancy between the design (initial) composition and the actual composition.

The study of the intrinsic properties of materials requires the use of quality controlled samples, whose phases and compositions have been identified from a solid state chemistry point of view.
For this purpose, it is essential to determine the actual composition of the sample after synthesis by compositional analysis, even if the sample is single phase.
As mentioned in section \ref{sec_intro} and discussed in more detail in section \ref{sec_Tdependence}, the dopant distribution in M-type ferrite can be strongly dependent on the cooling conditions or post-annealing of the sample, even if the compositions of the samples are the same.
This means that the heat treatment conditions should be carefully controlled so that they are the same when comparing the properties of different samples.
The necessary conditions for the sample used in this study were as follows: (i) a state close to equilibrium was achieved by properly controlling the heat treatment conditions, (ii) it was single phase (no secondary phases), (iii) the composition was uniform throughout the sample, and (iv) the actual composition was determined.
To satisfy these conditions, we used single crystal samples grown by the flux method using a platinum crucible, which is a popular growth method for M-phase single crystals \cite{Gambino1961}.
To satisfy (i), the sample was slowly cooled at a predetermined cooling rate. 
With natural cooling, it was difficult to standardise the cooling conditions for each sample.
The samples were grown in air (oxygen partial pressure 0.2 atm) or in oxygen gas flow (1 atm);
these atmospheres are easy to handle and allow stable crystal growth, facilitating (i)--(iii).
However,  Fe$^{2+}$ contamination is inevitable and care must be taken to ensure that the effect of Fe$^{2+}$ does not vary significantly from sample to sample. 
In the following NMR experiments, powdered crystals were used; the use of single crystals as raw material guarantees (ii).
Although other crystal growth methods, such as the travelling solvent floating zone (TS-FZ) method, are known to grow large single crystals \cite{Ueda2017}, crystals grown in a crucible were used in this study.
This is because, while the FZ method facilitates compositional control by growing crystals from the melt in a closed space isolated in air, it limits the freedom of atomic flow in the melt, and the sample is generally cooled rapidly, which may cause the synthesised crystals to deviate from their equilibrium state.
In the crucible flux method, the crystals grow gently in the melt without significant perturbation, resulting in the growth of crystals in a state close to equilibrium (note that this is not necessarily the ideal equilibrium phase, as described in section \ref{sec_intro}), and (i) and (iii) are satisfied.
Indeed, $^{59}$Co-NMR spectra of La--Co co-substituted SrFe$_{12}$O$_{19}$ show slightly different characteristics for crystals synthesised by the TS-FZ method and crystals synthesised by the flux method using a crucible \cite{Sakai2018}.

Wavelength dispersive X-ray spectrometry (WDX) with a scanning electron microscope (SEM) was used to satisfy condition (iv).
Unlike energy dispersive X-ray analysis (EDX), WDX can completely separate the Fe and Co spectra, providing a highly accurate analysis of Co composition.
X-ray analysis, which only observes the sample surface, has been criticised for its low accuracy.
However, comparison with inductively coupled plasma atomic emission spectrometry (ICP-AES), which gives the average value for the bulk sample, shows that the two analyses are in good agreement \cite{Ueda2017, Uji2017}.
The high reliability of WDX has been demonstrated in many studies \cite{Shimoda2016, Waki2023a, Uji2017, Waki2019, Waki2020a, Waki2020b, Kassem2022, Waki2023b, Nakai2023, Waki2023c}.
In addition, WDX, which analyses a narrow spot where X-rays are irradiated, guarantees (iii) by sampling different parts of the crystal.

We have applied this sample strategy to systematic studies of La substituted and La--Co co-substituted SrFe$_{12}$O$_{19}$ and CaFe$_{12}$O$_{19}$.
Their structural and magnetic properties have been reported in detail in \cite{Shimoda2016, Uji2017, Waki2019}.
In this study, among the single crystals synthesised in these series of studies, we selected those with a Co composition close to 0.2 and in sufficient quantity
(see table \ref{table_sample} and  section \ref{sec_sample} for details).
Not all samples had a Co composition of exactly 0.2. 
This is because the final composition of flux-grown M-phase crystals is difficult to deliberately control; the M-phase melts incongruently \cite{Langhof2009}, and the elemental fraction of the reagent before melting and the composition of the grown crystal are generally different.
Nevertheless, the compositions determined by WDX were sufficiently accurate and the samples used in this study satisfy conditions (i)--(iv) above.
It should be noted that the information used in this study is the ratio of Co occupancy among sites, not the absolute amount, so that differences in Co composition within reasonable limits will not affect the outcome of the discussion.
As the samples used in this study were synthesised at relatively low oxygen potentials, the formation of Fe$^{2+}$ was inevitable.
However, since the design concept of La--Co co-substitution is to prevent the formation of Fe$^{2+}$, and the purpose of this study was to investigate the site-to-site distribution of divalent Co, we assumed that the effect of Fe$^{2+}$ in the samples used in this study could be neglected.
In fact, the effect of Fe$^{2+}$ on the magnetic properties is not significant in the present samples; the presence of Fe$^{2+}$ reduces the uniaxial magnetic anisotropy in regions of low Fe$^{2+}$ concentration \cite{Ueda2017, Kupferling2006}.
The physics of Fe$^{2+}$ in M-type ferrite is certainly important \cite{Lotgering1974, Chlan2015}, but is not considered in this study; the discussion is limited to outside its influence.
In La substituted CaFe$_{12}$O$_{19}$, the Fe/(Ca + La) content ratio is known to be less than 12 \cite{Lotgering1980}.
Some Ca may occupy Fe sites, but the origin of this anomaly has not been fully understood \cite{Kobayashi2017, Waki2019, Pollert1985, Blanco1991}.
We will not address this point in detail in this study.

\subsubsection{Evaluation method for site-to-site distribution of Co}
\label{sec_NMRstr}

What we need in this study is a rapid method for assessing the site-to-site distribution of Co.
The current understanding of the Co occupancy sites in La--Co co-substituted SrFe$_{12}$O$_{19}$ and CaFe$_{12}$O$_{19}$ was first achieved by combining synchrotron radiation with neutron experiments \cite{Kobayashi2011, Kobayashi2016, Kobayashi2016T, Kobayashi2017}.
Experiments requiring large-scale facilities are certainly useful, but lack immediacy and availability.
Rapid on-site evaluation of synthesised samples is needed to accelerate research.

Nuclear methods, such as M\"ossbauer spectroscopy and NMR, have been widely used to assess cation distribution with high immediacy and availability.
However, their application has historically caused confusion in the identification of Co sites in La--Co co-substituted M-type SrFe$_{12}$O$_{19}$ \cite{Morel2005}.
Recently, a series of studies have been carried out to investigate the validity and interpretation of $^{57}$Fe M\"ossbauer spectroscopy and $^{57}$Fe, $^{59}$Co-NMR in this research area \cite{Sakai2018, Oura2018, Nagasawa2020, Nakamura2016}.
As a result, we have confirmed that although the $^{57}$Fe probe methods are useful for discussing microscopic magnetism, for example through experiments under a magnetic field, they are not ideal for quantifying the Co distribution.
In the case of M-type ferrite, quantitative analysis of the Co distribution requires the detection of trivial intensity changes in the $^{57}$Fe spectrum, which is composed of overlapping multi-components.
Information on Co is obtained from changes in the number of Fe atoms, but the effect of Co substitution is attenuated because Co is distributed over multiple sites.
The intensity of the $^{57}$Fe signal is more sensitive to the substitution modified environment than to the number of Fe atoms occupying the site \cite{Sakai2018, Oura2018}.
In addition, the lattice anharmonicity of M-type ferrites \cite{Rensen1969, Obradors1985, Obradors1988, Wang2014} makes it difficult to evaluate the true signal intensity and challenges the assumption that the signal intensity is proportional to the abundance of the corresponding nuclei.
Thus, the shortcomings of the $^{57}$Fe probe methods are more pronounced in the quantitative evaluation of the Co distribution.
This is in contrast to the fact that the $^{57}$Fe probe experiment works well for discussing more dramatic changes such as the crystal structure transition in M-type ferrites \cite{Kupferling2006, Chlan2015}.

In $^{59}$Co-NMR, on the other hand, the integrated intensity of the signal corresponds directly to the number of Co atoms occupying a particular site.
In principle, therefore, $^{59}$Co-NMR provides simple and direct information on the Co occupancy sites and substitution amounts, provided that the site assignment of the signal is complete. 
The NMR signal intensity of $^{59}$Co nuclei is generally extremely strong, and especially in ferromagnetic materials the signal intensity is greatly enhanced; thus, significant information can be extracted, even with a small amount of Co. 
The problem in previous studies of $^{59}$Co-NMR on La--Co co-substituted M-type ferrites \cite{Nakamura2016, Pieper2002, Kouril2013} was the site assignment of the signal and possibly the sample used.
The site assignment of the $^{59}$Co-NMR signal of La--Co co-substituted SrFe$_{12}$O$_{19}$ has already been completed with respect to the distinction between tetrahedrally and octahedrally coordinated sites (see section 2.1 in \cite{Nakamura2019}).
We believe that this conclusion is also valid for other La--Co co-substituted M-type ferrites.

In this study, the NMR spectra were recorded on a home-built state-of-the-art  broadband spectrometer at the Institut de Physique et Chimie des Mat\'eriaux de Strasbourg (IPCMS) using an untuned (high-pass design) probe \cite{Meny2021, Panissod2020}, which ensures good reproducibility of the frequency spectrum. 
This system is particularly suitable for the measurement of ferromagnetic (and ferrimagnetic) materials, whose resonance signals are distributed over a wide frequency range.
When there are multiple magnetic sites in different local environments, as in M-type ferrites, the local anisotropy of each site may be different.
In such cases, the signal enhancement factor will be different at each site, and it is necessary to correct for the frequency dependence of the enhancement factor in order to evaluate the distribution of nuclei corresponding to the resonance frequency.
The enhancement factor at each frequency point was evaluated by measuring the signal intensity at each frequency using multiple radiofrequency (RF) field powers.
The RF power dependence of the signal intensity was fitted with a suitable function, the width of which was also taken into account.
From the above series of procedures, the enhancement factor was used to correct the signal intensity.
Monitoring the enhancement factor also facilitates the differentiation of NMR signals from the magnetic domains and walls.
Intensities were corrected by dividing by the square of the frequency, which is standard in NMR methods.
The NMR spectrum corrected by the above process corresponds to the distribution of nuclei corresponding to the resonance frequencies and is independent of the value and distribution of the enhancement factor.
Furthermore, it is easy to calculate the local magnetic anisotropy of each environment from the signal enhancement factor, providing a unique insight into the magnetic properties of the sample, which has been described in detail in \cite{Meny2021, Panissod2020}.

Note that the signal intensity is corrected by dividing it by the square of the frequency and that the resonance frequency for our materials is significantly spread over the frequency domain (the lower and upper frequency limits differ by an order of magnitude, see section \ref{sec_NMR}).
This means that the correction factor differs by a factor of $10^2$ between the lowest and highest frequencies.
Thus, if the correction factors contain potential errors, however small, these errors will be amplified in the final intensities after correction.
Therefore, the final quantitative value of the Co distribution may contain an error of the order of 10\%.
Nevertheless, the quantitative nature of the system at IPCMS is exceptional for this type of experiment.
Moreover, the errors contained in the correction coefficients are expected to vary systematically with frequency, so that the qualitative trend of the intensity distribution cannot be seriously affected.
Furthermore, it should be noted that even sophisticated EXAFS and XMCD measurements show overlapping spectra of several components derived from Co occupying several sites.
In order to quantitatively assess the amount of Co at each site, the intensity of each component must be decomposed by fitting.
In contrast, in the $^{59}$Co-NMR experiment, the signal derived from Co occupying each site was observed without overlap (see figure \ref{fig_NMR}).
In short, the direct quantification, speed and simplicity of the $^{59}$Co-NMR method are outstanding among experimental techniques for the same purpose.
It is therefore undeniable that the $^{59}$Co-NMR is the most appropriate method for this study.
This study follows a series of studies in which $^{59}$Co-NMR experiments have been applied to La--Co co-substituted M-type ferrites from the above perspective \cite{Nakamura2019, Nakamura2016}.

In this study single crystals were powdered for NMR measurements.
As explained in section \ref{sec_samplestr}, the reason for using single crystals is to ensure that the sample is single phase.
As the main objective of this study was to assess the Co occupancy between sites, single crystals were powdered.
Sakai \textit{et al} \cite{Sakai2018} reported a difference in NMR results between powdered and single crystal samples for the same sample type; there is anisotropy in the excitation of Co nuclei in this system, and this anisotropy is site dependent \cite{Nakamura2019}.
This complicates the process of evaluating Co occupancy ratios from single crystal NMR spectra.
In particular, the fact that the 12k site has low local symmetry and multiple local symmetry axes further complicates the analysis.
In this study, we have avoided this problem by powdering the crystals and measuring the average intensity.

\subsubsection{Prediction of Co site preference}
\label{sec_DFTstr}

First-principles total energy calculations based on DFT have been used to help interpret the experimental results.
The subject of discussion is the site-to-site distribution of Co, that is, atomic-level issues.
One of the objectives is also to perform structural optimisation calculations to discuss the stable structure of the crystal.
Therefore, the Vienna Ab initio Simulation Package (VASP) \cite{Kresse1993, Kresse1994, Kresse1996a, Kresse1996b} with a plane-wave basis and the projector-augmented wave (PAW) method \cite{Bloch1994, Krasse1999}, which is suitable for total energy calculations with fast structure optimisation, was chosen as the calculation code.
The use of the PAW method reduces the computational load.
However, the PAW method does not explicitly treat the core orbitals.
Therefore, VASP is less effective in discussing phenomena where the contribution of the core electrons is dominant, such as hyperfine interactions at transition metal sites.
Therefore, this DFT calculation is only used to predict the Co site preference.
The physics of the anharmonic lattice vibrations of Fe at the 2b site \cite{Rensen1969, Obradors1985, Obradors1988, Wang2014} was also not investigated in depth in this study.
Anharmonic fluctuations at the 2b site are likely to be intrinsically important for the multiferroic nature of hexagonal ferrites \cite{Wang2014, Kimura2012}.
However, we assume that the effect of Co substitution on the anharmonicity of the 2b site is not pronounced, and the possibility of Co substitution at the 2b site has been experimentally ruled out \cite{Kobayashi2011, Kobayashi2016, Kobayashi2016T, Kobayashi2017}.

Previous DFT studies have attempted to predict Co occupancy sites in La--Co co-substituted SrFe$_{12}$O$_{19}$ \cite{Ohtsuka2016, Hui2014}.
Hui \textit{et al} \cite{Hui2014} predicted that the priority occupancy site of Co is the 2a site in their calculations using generalised gradient approximations (GGA) \cite{Perdew1996, Perdew1997}.
This differs from the current view.
Ohtsuka \textit{et al} \cite{Ohtsuka2016} performed calculations using GGA with Hubbard-type orbital interaction  (GGA+$U$), but failed to reproduce the experimental results of the Co occupancy site.
Calculations using GGA+$U$ are relatively expensive, but require the provision of appropriate Hubbard $U$ values for the transition metal atoms.
However, in addition to the fact that the appropriate $U$ is not clear a priori, it has already been pointed out that DFT calculation results for M-type ferrites are sensitive to $U$ \cite{Chlan2015, Wu2016A, Tejera2021}.
Our preliminary calculations using GGA+$U$ showed that even if a particular $U$ could explain the properties of a particular substance, the same $U$ could not uniformly explain the overall trend among different substances.
Therefore, we will attempt a more accurate calculation that is feasible on realistic time scales and does not include arbitrary parameters, such as $U$.
For this purpose, we have chosen the Heid-Scuseria-Ernzerhof hybrid exchange-correlation functional (HSE06) \cite{Heyd2003}.
To facilitate the convergence of the calculations, we will first perform a structure optimisation with the GGA+$U$ calculation using the appropriate $U$ and use the resulting structure parameters as initial values for the HSE06 calculation.

Calculations using HSE06 are relatively computationally intensive, and in order to perform the calculations in a realistic time, the structural model must be suitably simplified.
The model for the Co substituent of $A$Fe$_{12}$O$_{19}$ is described below.
First, the interaction between different Co atoms was neglected, assuming that the Co atoms were sufficiently dilute.
This is because considering the interactions between Co atoms requires considering many cases with different arrangements of neighbouring Co atoms.
We consider a supercell of size $2 \times 2 \times 1$ times the unit cell as the smallest supercell that satisfies the condition that Co is sufficiently dilute and replaces one of the $24 \times 4 = 96$ Fe atoms with a Co atom.
This corresponds to a substitution of 0.125 Co for a Fe+Co composition of 12.
This Co composition is lower than that of the samples used in the NMR experiment ($\sim$0.2).
However, it was assumed that the condition where Co is sufficiently diluted is also fulfilled in the experiment, and a comparison was made between the calculations and the experiment.
Supercells of the five types, with Co substituted at each Fe site, were studied.

Second, the effect of La substitution was assumed to be replaced by the effect of one additional electron per supercell.
In real materials, La$^{3+}$ is replaced by $A^{2+}$ at the same time as the Co substitution for charge compensation.
This means that Fe$^{3+}$ in $A$Fe$_{12}$O$_{19}$ is replaced by Co$^{2+}$ instead of Co$^{3+}$.
However, in models with the co-substitution of La and Co, there are multiple relative positions of Co and La, and the number of such cases is large.
This assumption is equivalent to neglecting the interaction between La and Co, the model seems reasonable as a first approximation.
The following calculations confirm that the value of the Co bond valence sum is in the range of +1.9 to +2.3 in all cases (except for 4f$_2$-Co for $A$ = Ca, see section \ref{sec_DFT_CaBa}).
The magnitude of the magnetic moment at this time was approximately 2.6 $\mu_\mathrm{B}$/Co.
In other words, Co was converging to a high spin state, close to the divalent state.
This suggests that the structural model is working to some extent.
In fact, it is quite possible that the partial substitution of the $A$ site by La affects the electronic state of the system and the priority occupancy site of Co.
To confirm this, it is necessary to perform structural optimisation calculations for systems with both La and Co substitutions; 
this will be reserved for future research.

\begin{table}[t]
\caption{\label{table_sample}
\noindent
List of samples used for $^{59}$Co-NMR measurements, their lattice constants at room temperature and their anisotropy fields at 5 K.}
\small
\begin{tabular}{cccccccc}
\br
Label & Chemical formula & $r(A^{2+})$$^\dag$ & $a$ & $c$ & $c/a$ & $H_\mathrm{A}$ &Ref.\\  
& & (\AA) &  (\AA)  &  (\AA)  &  & (T) &\\
\mr
Ca17 & Ca$_{0.77}$La$_{0.37}$Fe$_{11.70}$Co$_{0.17}$O$_{19}$ & 1.34 & 5.8890 & 22.919 & 3.8918 & 2.689 & \cite{Waki2019}\\
Sr28 & Sr$_{0.68}$La$_{0.32}$Fe$_{11.72}$Co$_{0.28}$O$_{19}$ & 1.44  & 5.8839 & 23.005 &3.9098 & 2.540 &\cite{Nakamura2019} \\
Ba20 & Ba$_{0.70}$La$_{0.30}$Fe$_{11.80}$Co$_{0.20}$O$_{19}$ & 1.61 & 5.8910 & 23.141 & 3.9282 & 2.083 & --\\
\br
\end{tabular}\\
$^\dag$ 12-coordinated ionic radius \cite{Shannon1976}.\\
\end{table}

\subsection{Sample preparation and characterization}
\label{sec_sample}

The samples used for the $^{59}$Co-NMR experiments were selected from our stock of crystals, one each for $A$ = Ca, Sr and Ba in La--Co co-substituted $A$Fe$_{12}$O$_{19}$, following the strategy described in section \ref{sec_samplestr}.
Samples with Co compositions around 0.2 were selected (see table \ref{table_sample}).
These three samples are referred to as Ca17, Sr28 and Ba20, respectively.
For comparison of the $A$-ion size, the ionic radius of the 12-coordination by Shannon \cite{Shannon1976} is added to table \ref{table_sample}.

Crystal growth techniques for Ca17 and Sr28 have been reported in \cite{Waki2019, Nakamura2019}.
Systematic studies of flux-grown single crystals of La--Co co-substituted SrFe$_{12}$O$_{19}$ and CaFe$_{12}$O$_{19}$ have been reported in detail \cite{Shimoda2016, Waki2019}.
The Ba20 crystals were grown under the same conditions as Ca17.
All crystals were cooled at $-5$\,$^\circ$C/h above 1000\,$^\circ$C and $-200$\,$^\circ$C/h below 1000\,$^\circ$C to ensure the same cooling conditions. 
All compositions shown in table \ref{table_sample} are actual compositions determined for the grown crystals using SEM-WDX (INCA wave 500, Oxford Instruments attached to SEM S-3500H, Hitachi).
For Sr28 and Ba20, the (Fe+Co)/($A$+La) composition ratio was 12, whereas for Ca17, the (Fe+Co)/(Ca+La) ratio was greater than 12, which is characteristic of CaFe$_{12}$O$_{19}$ as mentioned in section \ref{sec_samplestr}.
The room temperature lattice constants  given in table \ref{table_sample} \cite{Nakamura2019, Waki2019} were estimated by X-ray diffraction (XRD) using a PANalytical X'Pert PRO Alpha-1 for partially powdered single crystals.

The anisotropy field $H_\mathrm{A}$ at 5 K shown in table \ref{table_sample} \cite{Nakamura2019, Waki2019} was obtained from the analysis of single crystal magnetisation curves, which were measured with a SQUID magnetometer (Quantum Design, MPMS).
The anisotropy energy $E_\mathrm{A}$ was evaluated from the region bounded by the easy and hard axis magnetisation curves.
Using the values of $E_\mathrm{A}$ and the saturation magnetisation $M_\mathrm{s}$, $H_\mathrm{A}$ was obtained from the relation $H_\mathrm{A} = 2E_\mathrm{A}/M_\mathrm{s}$.
The results of our systematic magnetic measurements of La--Co co-substituted CaFe$_{12}$O$_{19}$ and SrFe$_{12}$O$_{19}$ have been reported previously \cite{Shimoda2016, Waki2019}.
The magnetisation curve of single crystalline BaFe$_{12}$O$_{19}$ \cite{Casimir1959} is qualitatively the same as that of SrFe$_{12}$O$_{19}$, except that the magnetic field at which the hard axis magnetisation reaches saturation (i.e., $H_\mathrm{A}$) is slightly lower.
 In the B20 material used in this study, the effect of La--Co co-substitution on BaFe$_{12}$O$_{19}$ is qualitatively the same as that on SrFe$_{12}$O$_{19}$.
Although the $M_\mathrm{s}$ values of the three samples were almost the same, the absolute value of $M_\mathrm{s}$ is not discussed in this paper for the following reasons.
Taking into account the Co distribution revealed in section \ref{sec_NMR}, the change in magnetisation with Co substitution in this system is expected to be at most 0.5\%.
Co substitution makes the magnetisation curve convex, and suppresses the saturation of magnetisation even under a magnetic field of 7 T \cite{Shimoda2016, Waki2019}.
In addition, when measuring the hard axis magnetisation of single crystals, the sample may tilt slightly during the measurement due to the large torque exerted on the sample by the magnetic field.
Therefore, it is technically difficult to experimentally detect the expected change in $M_\mathrm{s}$ due to Co substitution, and even if such a change is observed, it is not appropriate to simply discuss its origin.

\subsection{NMR}
\label{sec_NMRexp}

The $^{59}$Co-NMR experiments were performed at IPCMS, following the strategy described in section \ref{sec_NMRstr}.
The nuclear quantities for $^{59}$Co are the nuclear spin $I = 7/2$, the gyromagnetic ratio $\gamma /2\pi = 10.05$ MHz T$^{-1}$, the quadrupole moment $Q = 42.0 \times 10^{-30}$ m$^2$ and a natural abundance of 100\%.
For the reasons described in section \ref{sec_NMRstr}, the single crystals were powdered for NMR measurements.
Measurements were performed at 2 K under zero external field in the frequency range 40--600 MHz using 1.28 MHz steps. 
A classical spin-echo sequence of RF pulses was used. 
To observe magnetic domain excitations, we used wide pulses (12.8 $\mu$s) separated by a 5 $\mu$s delay. 
The maximum RF power range of our setup was used ($\le$ 32--40 dBm). 
Each spectrum results in 5,000 to 10,000 frequency scans, depending on the mass of the sample (100 to 200 mg). 
To evaluate the signal enhancement factor, the signal intensity was measured as a function of RF power (four or more points) at each frequency.
The NMR data for Sr28 have already been reported in \cite{Nakamura2019} and some of the results of this study are quoted from them.

\subsection{DFT calculations}
\label{sec_DFTmethod}

The stable structures and total energies of the undoped materials and Co substituents of $A$Fe$_{12}$O$_{19}$ ($A$ = Ca, Sr and Ba) were evaluated by structure-optimisation calculations using VASP.
The convergence condition was set to $1 \times 10^{-5}$ eV ($1 \times 10^{-4}$ eV for GGA+$U$ preliminary calculations) for the electronic self-consistent loop and 0.01 eV/\AA\ for the structure optimisation.
The $k$-mesh was $2 \times 2 \times 1$ and the energy cutoff was 500 eV.
The experimental values reported in \cite{Obradors1988, Nemrava2017} were used as the initial values of the structural parameters in the GGA+$U$ calculations. 
However, since CaFe$_{12}$O$_{19}$ does not exist in the equilibrium phase, the value of SrFe$_{12}$O$_{19}$ was used for the calculations when $A$ = Ca.

Calculations for undoped $A$Fe$_{12}$O$_{19}$ were carried out using the following procedure, assuming that the Fe sublattice has a conventional ferrimagnetic structure collinear along the $c$-axis.
(i) Preliminary GGA+$U$ calculations using arbitrary values in the range 4 to 6 eV as $U$ for Fe.
(ii) HSE06 calculations with the structural parameters optimised in (i) as initial values (the results are treated as final). 
(iii) Optimal GGA+$U$ calculations using the $U$ value ($U$(Fe) = 4.1 eV) obtained by the method described at the end of this section (the results are treated as reference data).

\renewcommand{\vec}[1]{{\mbox{\boldmath $#1$}}}

For the Co substituents, as described in section \ref{sec_DFTstr}, the $2 \times 2 \times 1$ supercell of the $A$Fe$_{12}$O$_{19}$ unit cell was considered, in which one of the Fe ions was replaced by Co$^{2+}$ (valence initially set to 2),  and then an electron was added.
Supercells of the five types, with Co substituted at each Fe site, were studied.
The magnetic moments $\vec{m}$ of the host Fe and the doped Co were assumed to be collinear along the $c$-axis.
In this case, $\vec{m}$(Co) can be parallel or antiparallel to $\vec{m}$(Fe).
We calculated both cases for $A$ = Sr, but only the parallel case for $A$ = Ca and Ba, considering the results for $A$ = Sr (section \ref{sec_Mdirection}).
The calculations were carried out as follows:
(i) Preliminary GGA+$U$ calculations with $U$(Fe) = 6 eV and $U$(Co) = 5 eV.
(ii) HSE06 calculations with the structural parameters obtained in (i) as initial values (the results are treated as final).

The optimal $U$ values in the GGA+$U$ calculations were sought to reproduce the total energy difference given by the HSE06 calculations for the Co substituents of SrFe$_{12}$O$_{19}$.
Under the constraint $U$(Fe)/$U$(Co) = 1.2, the HSE06 results are best reproduced when $U$(Fe) = 4.1 eV and $U$(Co) = 3.4 eV.
Based on this result, the structural parameters of undoped $A$Fe$_{12}$O$_{19}$ were again optimised by GGA+$U$ calculations assuming $U$(Fe) = 4.1 eV (reference data in tables \ref{table_latticeconst} and \ref{table_volume}).
However, the GGA+$U$ calculations for the Co substituents of CaFe$_{12}$O$_{19}$ using the same $U$ values failed to reproduce the experimental tendency of the Co distribution (section \ref{sec_NMR}); 
the calculations predicted that 12k-Co and 2a-Co are more stable than 4f$_1$-Co in CaFe$_{12}$O$_{19}$.

\section{Results}

\subsection{$A$-ion dependence of structural and magnetic parameters}
\label{sec_parameter}

The lattice constants of the three samples measured at room temperature and the anisotropy field,  $H_\mathrm{A}$, at 5 K are listed in table \ref{table_sample}.
The values for Ca17 and Sr28 are taken from \cite{Shimoda2016, Waki2019}.
The parameter $a$ changes little and the difference in $a$ is at most 0.1\%.
In contrast, $c$ is highest for $A$ = Ba and lowest for $A$ = Ca with a difference of approximately 1\%.
This means that the effect of the $A$-ion size is anisotropic.
This may be because $a$ is determined by the skeleton of the oxygen ions, whereas $c$ (particularly the R-block length) directly reflects the effect of the $A$-ion size.
Thus, the change in $A$ ions acts as a uniaxial strain.
In the following, $c/a$ is used as a parameter to characterise the structure of the three samples.

As shown in table \ref{table_sample}, the anisotropic field $H_\mathrm{A}$ (at 5 K) is negatively correlated with the $A$-ion size and the value of $c/a$.

\subsection{NMR}
\label{sec_NMR}

\begin{figure}[t]
\begin{center}
\includegraphics[width=0.45\textwidth]{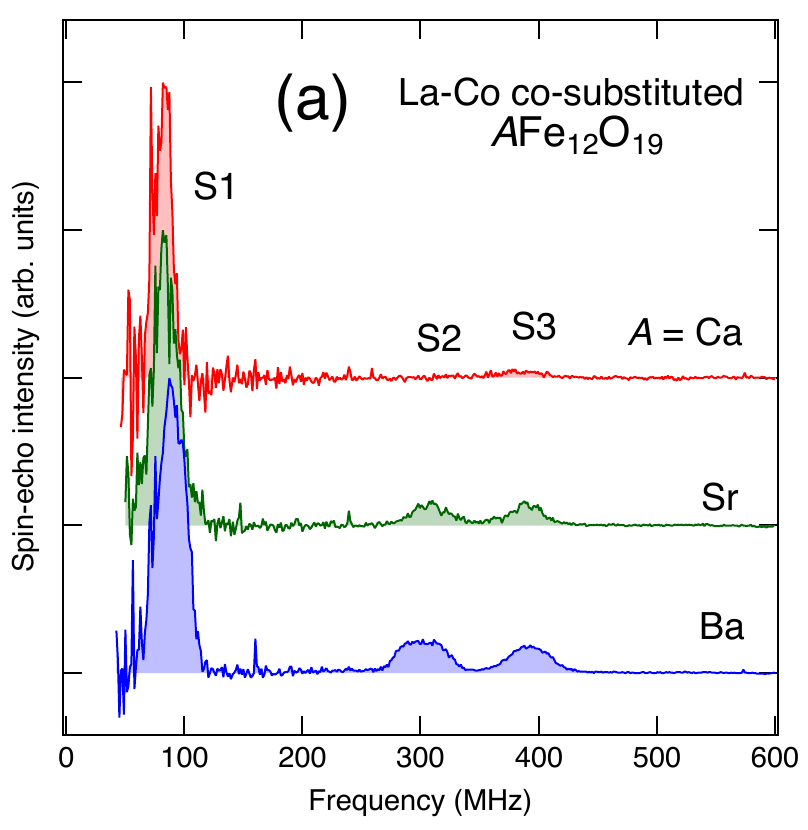}~~~
\includegraphics[width=0.3\textwidth]{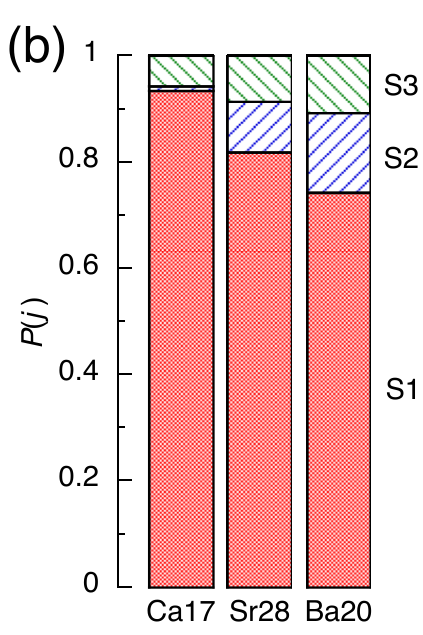}
\caption{\label{fig_NMR}
(a) $^{59}$Co-NMR spectra for La--Co co-substituted $A$Fe$_{12}$O$_{19}$ measured at zero applied field and at 2 K. 
The intensity was normalised so that the maximum intensity of S1 is the same.
(b) Intensity ratios of S1, S2 and S3 for the three samples.
}
\end{center}
\end{figure}

Figure \ref{fig_NMR}(a) shows the $^{59}$Co-NMR spectra of the three samples measured at zero field and 2 K.
Three types of signals, S1, S2 and S3, were observed in all samples as in the previous studies \cite{Nakamura2019, Nakamura2016}. 
An additional signal (S4) was observed at $\sim$530 MHz in Sr28 \cite{Nakamura2019}.
The origin of S4 has been attributed to high spin Co$^{3+}$, which may be present in very small amounts due to an unexpected charge imbalance.
However, its intensity is less than 0.2\% of the total Co signal intensity and is therefore ignored in this study.
No S4 signals of significant intensity were observed for Ca17 and Ba20.
Although the signal intensity ratios vary from sample to sample, it is noteworthy that three similarly shaped signals were observed in almost all common frequency ranges.
This indirectly suggests that various parameters other than the differences in $A$ ions remain nearly constant during crystal growth.
The intensity of S1 was the highest for all samples.
In figure \ref{fig_NMR}(a) the signal intensity is normalised to the intensity of S1.
The centre frequencies of S1, S2 and S3, evaluated by fitting to a Gaussian function, are listed in table \ref{table_NMR}.
They differed slightly from sample to sample.
This frequency shift is discussed in the last part of section \ref{sec_localdistortion}.

Figure \ref{fig_NMR}(a) shows that the intensities of S2 and S3 tend to decrease with decreasing $A$-ion size ($A$ = Ba $\rightarrow$ Sr $\rightarrow$ Ca).
The proportions $P(j)$ ($j$ = S1, S2, S3) of the integrated intensities of S1, S2 and S3 for each sample are shown in figure \ref{fig_NMR}(b).
As the $A$-ion size decreased, $P(\mathrm{S1})$ increased.
$P(\mathrm{S3})$ decreased slightly and the change was small.
In contrast, $P(\mathrm{S2})$ decreased significantly and at $A$ = Ca it decreased to approximately 1.5\% of the total.
That is, although S2 and S3 were observed in similar frequency ranges, the $A$-ion dependence of their intensities was large for $P(\mathrm{S2})$ and small for $P(\mathrm{S3})$. 

The site assignments of the S1, S2 and S3 resonances when $A$ = Sr have been described in detail in section 3.1 of \cite{Nakamura2019}.
In this study, we apply this conclusion to the other $A$-ion cases as well; the S1 resonance is assigned to 4f$_1$-Co and the remaining S2/S3 are assigned to 2a/12k-Co.
Kobayashi \textit{et al} \cite{Kobayashi2016, Kobayashi2016T, Kobayashi2017} reported that Co occupies three sites: 2a, 4f$_1$ and 12k, with 4f$_1$ occupying the largest fraction, also for La--Co co-substituted CaFe$_{12}$O$_{19}$, according to studies using synchrotron radiation and neutrons.
This result supports our site assignment.
In \cite{Nakamura2019} we could not conclude whether S2 or S3 was assigned to 2a-Co or 12k-Co. 
However, considering the cation diffusion processes during the synthesis procedure, the occupancy of 12k-Co is expected to be sensitive to various conditions (including $A$-ion differences), whereas that of 2a-Co is relatively invariant (see section \ref{sec_Tdependence} for details).
The $A$-ion dependence of the signal intensity was large in S2 and small in S3 (figure \ref{fig_NMR}(b)), qualitatively suggesting an S2 $\rightarrow$ 12k and S3 $\rightarrow$ 2a assignment.

\begin{figure}[t]
\begin{center}
\includegraphics[width=0.8\textwidth]{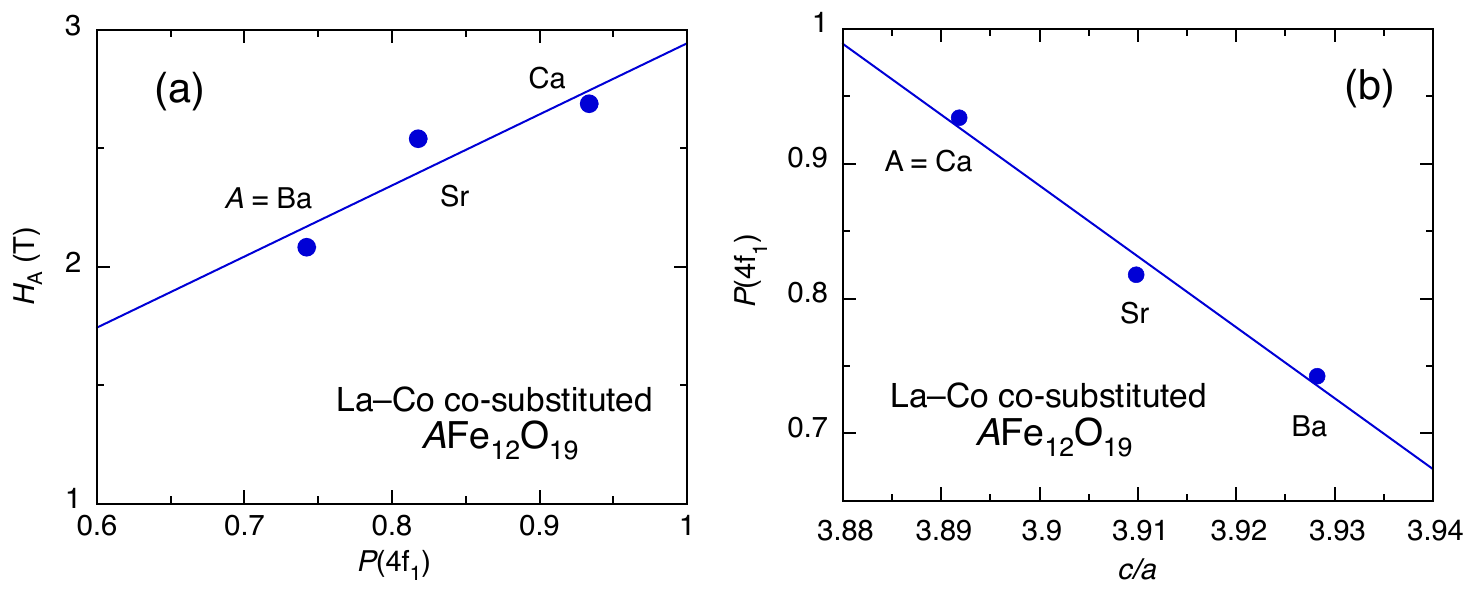}~~~
\caption{\label{fig_p(4f1)}
(a) Dependence of the anisotropy field on the 4f$_1$ site selectivity of Co. 
(b) $c/a$ dependence of the 4f$_1$ site selectivity of Co in La--Co co-substituted $A$Fe$_{12}$O$_{19}$ ($A$ = Ca, Sr and Ba) evaluated from the $^{59}$Co-NMR signal intensity.
The straight lines in (a) and (b) are visual aids.
}
\end{center}
\end{figure}

\begin{table}[t]
\caption{\label{table_NMR}
$^{59}$Co ferromagnetic nuclear resonance frequencies of La--Co co-substituted $A$Fe$_{12}$O$_{19}$ at zero field and at 2 K, corresponding internal fields and their site assignments.
Each $^{59}$Co-NMR line was fitted with a Gaussian function and the median was taken as the resonance frequency.
The direction of the local magnetic moment at each site and the direction of the internal field are defined as positive if they are parallel and negative if they are antiparallel. 
}
\begin{indented}
\item[]
\begin{tabular}{ccccccccc}
\br
Signal & \multicolumn{3}{c}{$\nu_\mathrm{res}$ (MHz)} && \multicolumn{3}{c}{$H_\mathrm{int}$ (T)} & Site\\  
&  Ca17  &  Sr28  &  Ba20 &&  Ca17  &  Sr28  &  Ba20 &\\
\mr
S1 & 82.4(4)  & 83.6(4) & 89.4(2) && $-8.20$(4) & $-8.32$(4) & $-8.90$(2) & 4f$_1$ \\
S2 & 327.0(29) & 308.3(6) & 301.3(4) &&  32.54(29) & 30.68(6)  & 29.98(4)  & 12k \\
S3 & 384.5(15)  & 389.1(5)  & 391.8(2) && 38.26(15) & 38.72(5) &  38.99(2) & 2a \\
\br
\end{tabular}\\
\end{indented}
\end{table}

Based on the site assignment above, the Co 4f$_1$ site occupancy $P$(4f$_1$) was highest at $A$ = Ca and lowest at $A$ = Ba.
Figure \ref{fig_p(4f1)}(a) shows the anisotropy field $H_\mathrm{A}$ of each sample at 5 K plotted against $P$(4f$_1$) (= $P$(S1)). 
These values are almost proportional to each other, indicating that the 4f$_1$ site selectivity of Co does indeed dominate the magnetic anisotropy.

As mentioned in section \ref{sec_parameter}, the difference in $A$ ions appears as a uniaxial strain.
Figure \ref{fig_p(4f1)}(b) shows the $c/a$ dependence of $P$(4f$_1$) for each sample.
$P$(4f$_1$) increased with decreasing $c/a$.
It is therefore hypothesised that the uniaxial compressive strain of the crystal improves the 4f$_1$ site selectivity of Co and consequently the uniaxial magnetic anisotropy.
As the 4f$_1$ site is a minority spin site, the magnetisation is simultaneously increased.
To quantitatively evaluate the effect of different Co compositions of the samples used in the experiments, one can refer to a previous study on La--Co co-substituted SrFe$_{12}$O$_{19}$ \cite{Nakamura2019}.
In this system, the change in $c/a$ was approximately 0.08\% when the Co concentration was changed by approximately 0.1.
Assuming that the Co occupancy ratio depends only on the value of $c/a$, the expected change in $P$(4f$_1$) does not exceed 0.02 (see figure \ref{fig_p(4f1)}(b)), which is sufficiently small compared with the change when the $A$ ion is replaced.
When $A$ = Ca, the Ca composition is greater than its stoichiometric composition.
If excess Ca occupies the 4f$_1$ site, as proposed by Kobayashi \textit{et al} \cite{Kobayashi2017}, the number of 4f$_1$ sites available for Co to occupy would be reduced, and hence $P$(4f$_1$) in Ca17 would be expected to be suppressed compared to Sr28 and Ba20.
However, this was actually increased.
This suggests that even if Ca partially occupies the 4f$_1$ site, the effect is insignificant and does not affect the discussion here.

\subsection{DFT calculations}

\begin{table}[t]
\caption{\label{table_latticeconst}
Lattice parameters, $c$-axis lengths of S and R blocks, and unit cell volumes of $A$Fe$_{12}$O$_{19}$ ($A$ = Ca, Sr and Ba) obtained by structure-optimisation DFT calculations with HSE06 and GGA+$U$ ($U$(Fe) = 4.1 eV).
Available experimental values at room temperature are also given.
No experimental values are available for $A$ = Ca because the undoped CaFe$_{12}$O$_{19}$ is not stable.
}
\begin{indented}
\item[]
\begin{tabular}{ccccccccc}
\br
$A$ & Method & $a$ (\AA) & $c$ (\AA) & $c/a$ & $c$(S) (\AA) & $c$(R) (\AA) & $V$ (\AA$^3$) & Ref.\\
\mr
Ca & HSE06 & 5.880 & 22.892 & 3.893 & 5.031 & 6.415 & 685.4\\
 & GGA+$U$ & 5.917 & 22.963 & 3.881 & 5.041 & 6.441 & 696.2\\
 & Experiment & -- & -- & -- & -- & -- \\
Sr & HSE06 & 5.885 & 23.017 & 3.912 & 5.025 & 6.483 & 690.4\\
 & GGA+$U$ & 5.931& 23.158 & 3.905 & 5.052 & 6.527 & 705.5\\
 & Experiment & 5.884 & 23.050 & 3.917 & 5.039 & 6.486 & 691.1 & \cite{Obradors1988} \\
Ba & HSE06 & 5.897 & 23.162 & 3.928 & 5.013 & 6.568 & 697.5\\
 & GGA+$U$ & 5.943 & 23.307 & 3.921 & 5.038 & 6.615 & 713.0\\
 & Experiment & 5.892 & 23.183 & 3.935 & 5.019 & 6.573 & 697.0& \cite{Nemrava2017} \\
\br
\end{tabular}\\
\end{indented}
\end{table}

\begin{table}[t]
\caption{\label{table_strparam}
Atomic coordinates of $A$Fe$_{12}$O$_{19}$ ($A$ = Ca, Sr and Ba) obtained by structure-optimisation DFT calculations with HSE06.
Experimental data for SrFe$_{12}$O$_{19}$ and BaFe$_{12}$O$_{19}$ were taken from \cite{Obradors1988} and  \cite{Nemrava2017}, respectively.
}
\begin{indented}
\item[]
\begin{tabular}{clllllllll}
\br
Element/Site & \multicolumn{3}{c}{HSE06} & \multicolumn{3}{c}{GGA+$U$} & \multicolumn{3}{c}{Experiment} \\
 & $x$ & $y$ &$z$ & $x$ & $y$ &$z$ & $x$ & $y$ &$z$ \\
\mr
CaFe$_{12}$O$_{19}$\\
Ca/2d & 0.6667 & 0.3333 & 0.2500  & 0.6667 & 0.3333 & 0.2500 & -- & -- & --\\
Fe/2a & 0.0000 & 0.0000 & 0.0000 & 0.0000 & 0.0000 & 0.0000 & -- & -- & --\\
Fe/2b & 0.0000 & 0.0000 & 0.2500 & 0.0000 & 0.0000 & 0.2500 & -- & -- & --\\
Fe/4f$_1$ & 0.3333 & 0.6667 & 0.0270 & 0.3333 & 0.6667 & 0.0275 & -- & -- & --\\
Fe/4f$_2$ & 0.3333 & 0.6667 & 0.1913 & 0.3333 & 0.6667 & 0.1915 & -- & -- & --\\
Fe/12k & 0.1691 & $0.3382$ & 0.8902 & 0.1688 & 0.3375 & 0.8902 & -- & -- & --\\
O/4e & 0.0000 & 0.0000 & 0.1521 & 0.0000 & 0.0000 & 0.1525 &-- & -- & --\\
O/4f & 0.6667 & 0.3333 & 0.0556 & 0.6667 & 0.3333 & 0.0557 & -- & -- & --\\
O/6h & 0.1817 & $0.3633$ & 0.2500 & 0.1821 & 0.3643 & 0.2500 & -- & -- & --\\
O/12k$_1$ & 0.1572 & $0.3144$ & 0.0527 & 0.1563 & 0.3125 & 0.0530 & -- & -- & --\\
O/12k$_2$ & 0.5056 & $0.0112$ & 0.1525 & 0.5061 & 0.0123 & 0.1523 & -- & -- & --\\
\mr
SrFe$_{12}$O$_{19}$\\
Sr/2d & 0.6667 & 0.3333 & 0.2500  & 0.6667 & 0.3333 & 0.2500 & $\frac{1}{3}$ & $\frac{2}{3}$ & $\frac{1}{4}$\\
Fe/2a & 0.0000 & 0.0000 & 0.0000 & 0.0000 & 0.0000 & 0.0000 & 0 & 0 & 0\\
Fe/2b & 0.0000 & 0.0000 & 0.2500 & 0.0000 & 0.0000 & 0.2500 & 0 & 0 & 0.2542\\
Fe/4f$_1$ & 0.3333 & 0.6667 & 0.0270 & 0.3333 & 0.6667 & 0.0271 & $\frac{1}{3}$ & $\frac{2}{3}$ & 0.0272\\
Fe/4f$_2$ & 0.3333 & 0.6667 & 0.1909 & 0.3333 & 0.6667 & 0.1910 & $\frac{1}{3}$ & $\frac{2}{3}$ & 0.1909\\
Fe/12k & 0.1689 & $0.3378$ & 0.8908 & 0.1688 & 0.3376 & 0.8909 & 0.1689 & $2x$ & 0.8907\\
O/4e & 0.0000 & 0.0000 & 0.1516 & 0.0000 & 0.0000 & 0.1520 & 0 & 0 & 0.1516\\
O/4f & 0.6667 & 0.3333 & 0.0552 & 0.6667 & 0.3333 & 0.0554 & $\frac{2}{3}$ & $\frac{1}{3}$ & 0.0552\\
O/6h & 0.1817 & $0.3634$ & 0.2500 & 0.1819 & 0.3638 & 0.2500 & 0.1817 & $2x$ & $\frac{1}{4}$\\
O/12k$_1$ & 0.1570 & $0.3141$ & 0.0524 & 0.1568 & 0.3136 & 0.0525 & 0.1565 & $2x$ & 0.0527\\
O/12k$_2$ & 0.5041 & $0.0083$ & 0.1511 & 0.5044 & 0.0088 & 0.1512 & 0.5047 & $2x$ & 0.1508\\
\mr
BaFe$_{12}$O$_{19}$\\
Ba/2d & 0.6667 & 0.3333 & 0.2500  & 0.6667 & 0.3333 & 0.2500 & $\frac{2}{3}$ & $\frac{1}{3}$ & $\frac{1}{4}$\\
Fe/2a & 0.0000 & 0.0000 & 0.0000 & 0.0000 & 0.0000 & 0.0000 & 0 & 0 & 0\\
Fe/2b & 0.0000 & 0.0000 & 0.2500 & 0.0000 & 0.0000 & 0.2500 & 0 & 0 & $\frac{1}{4}$\\
Fe/4f$_1$ & 0.3333 & 0.6667 & 0.0272 & 0.3333 & 0.6667 & 0.0272 & $\frac{1}{3}$ & $\frac{2}{3}$ & 0.0273\\
Fe/4f$_2$ & 0.3333 & 0.6667 & 0.1903 & 0.3333 & 0.6667 & 0.1903 & $\frac{1}{3}$ & $\frac{2}{3}$ & 0.1904\\
Fe/12k & 0.1687 & $0.3374$ & 0.8918 & 0.1686 & 0.3372 & 0.8919 & 0.1687 & $2x$ & 0.8917\\
O/4e & 0.0000 & 0.0000 & 0.1508 & 0.0000 & 0.0000 & 0.1511 & 0 & 0 & 0.1504\\
O/4f & 0.6667 & 0.3333 & 0.0555 & 0.6667 & 0.3333 & 0.0549 & $\frac{2}{3}$ & $\frac{1}{3}$ & 0.0548\\
O/6h & 0.1818 & $0.3635$ & 0.2500 & 0.1820 & 0.3640 & 0.2500 & 0.1821 & $2x$ & $\frac{1}{4}$\\
O/12k$_1$ & 0.1568 & $0.3136$ & 0.0519 & 0.1565 & 0.3130 & 0.0520 & 0.1563 & $2x$ & 0.0521\\
O/12k$_2$ & 0.5021 & $0.0042$ & 0.1491 & 0.5023 & 0.0046 & 0.1492 & 0.5024 & $2x$ & 0.1493\\
\br
\end{tabular}\\
\end{indented}
\end{table}

\begin{table}[t]
\caption{\label{table_volume}
Volume of the Fe--O coordination polyhedron of each Fe site in undoped $A$Fe$_{12}$O$_{19}$ ($A$ = Ca, Sr and Ba) 
calculated from atomic coordinates obtained by structure-optimisation DFT calculations with HSE06 and GGA+$U$ ($U$(Fe) = 4.1 eV) (table \ref{table_strparam}).
Experimental values calculated from the available atomic coordinates at room temperature are also given.
No experimental values are available for $A$ = Ca, as the undoped material CaFe$_{12}$O$_{19}$ is not stable.
}
\begin{indented}
\item[]
\begin{tabular}{cccccccc}
\br
$A$ & Method & \multicolumn{5}{c}{$V$ (\AA$^3$)} & Ref.\\
& & 2a & 2b & 4f$_1$ & 4f$_2$ & 12k \\
\mr
 Ca & HSE06 & 10.714 & 6.650 & 3.454 & 10.527 & 10.926\\
  & GGA+$U$ & 10.805 & 6.758 & 3.557 & 10.711 & 11.039\\
  & Experiment & -- & -- & -- & -- & --\\
 Sr & HSE06 & 10.693 & 6.730 & 3.461 & 10.646 & 10.946\\
  & GGA+$U$ & 10.922 & 6.863 & 3.559 & 10.871 & 11.187\\
  & Experiment & 10.706 & 6.737 & 3.498 & 10.730 & 10.893 & \cite{Obradors1988} \\
 Ba & HSE06 & 10.671 & 6.860 & 3.475 & 10.829 & 10.983\\
  & GGA+$U$ & 10.897 & 7.002 & 3.572 & 11.062 & 11.223\\
  & Experiment & 10.671 & 6.871 & 3.482 & 10.783 & 10.999 & \cite{Nemrava2017}\\
\br
\end{tabular}\\
\end{indented}
\end{table}

\subsubsection{Undoped $A$Fe$_{12}$O$_{19}$}

We have carried out structure-optimisation DFT calculations on $A$Fe$_{12}$O$_{19}$ to obtain information on the stable lattice constants and local structure of $A$Fe$_{12}$O$_{19}$ in the undoped form.
The lattice constants obtained from the calculations are given in table \ref{table_latticeconst}, the atomic coordinates in table \ref{table_strparam} and the volume $V$ of the Fe--O coordination polyhedron of each Fe site calculated from the atomic coordinates in table \ref{table_volume}.
The values obtained using GGA+$U$ are also listed for reference, together with calculations using HSE06.
The experimental values at room temperature \cite{Obradors1988, Nemrava2017} are also given for Sr and Ba ferrites. 
The S-block length $c$(S) and R-block length $c$(R) calculated for $A$Fe$_{12}$O$_{19}$ are also given in table \ref{table_latticeconst}, together with the $c$-axis length.
The sum of $c$(S) and $c$(R) is equal to $c/2$.
As shown in figure \ref{fig_crystal}, the 2b and 4f$_2$ coordinated polyhedra belong to the R block, while the 2a and 4f$_1$ coordinated polyhedra belong to the S block.
The 12k site is located at the boundary between the R and S blocks.
As a general trend, the experimental values in tables \ref{table_latticeconst}, \ref{table_strparam} and \ref{table_volume} are closer to the values calculated using HSE06 than those calculated using GGA+$U$.
This indicates that the calculations using HSE06 are more accurate than those using GGA+$U$.
In the following discussion only the values calculated using HSE06 are used.

\begin{table}[t]
\caption{\label{table_energy} Energy change evaluated by DFT calculations with HSE06 when one of the Fe atoms in the $A$Fe$_{12}$O$_{19}$ supercell is replaced by a Co atom, based on the energy when Co occupies the 4f$_1$  site and has a moment in the same direction as the host Fe.
}
\begin{indented}
\item[]
\begin{tabular}{ccrrrrr}
\br
$A$ &Moment$^a$ & \multicolumn{5}{c}{$\Delta E$ (meV)}\\
&& 2a & 2b & 4f$_1$ & 4f$_2$ & 12k\\
\mr
Ca$^b$ &$\uparrow \uparrow$ & 43 & 1628 & 0 & -- & 174\\
Sr &$\uparrow \uparrow$ & 30 & 840 & 0 & 818 & 139\\
Ba &$\uparrow \uparrow$ & 66 & 796 & 0 & 797 & 109\\
\mr
Sr &$\uparrow \downarrow$& 334 & 954 & 260 & 1222 & 362\\
\br
\end{tabular}\\
$^a$ Relative spin direction of substituted Co and host Fe.
$\uparrow \uparrow$ and $\uparrow \downarrow$ denote cases where the moment of the substituted Co is in the same or opposite direction to the host Fe, respectively. \\
$^b$ No stable solution with Co$^{2+}$ occupying the 4f$_2$ site was obtained in the case of $A$ = Ca (see text).
\end{indented}
\end{table}

\begin{figure}[t]
\begin{center}
\includegraphics[width=0.4\textwidth]{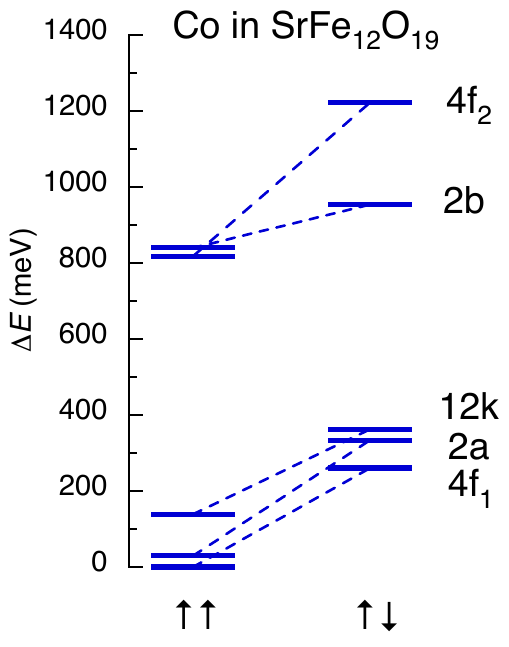}
\caption{\label{fig_Sr_dE}
Energy change evaluated by DFT calculations with HSE06 when each Fe site in SrFe$_{12}$O$_{19}$ is replaced by Co.
Both cases where the Co moment is in the same direction as and opposite to the magnetic moment of the host Fe are shown.
The energy of the case where Co occupies the 4f$_1$ site and the moment of Co is in the same direction as the magnetic moment of the host Fe is taken as the origin.
}
\end{center}
\end{figure}

\subsubsection{Co substituted SrFe$_{12}$O$_{19}$}
\label{sec_Mdirection}

The total energy $E(i)$ of the supercell, in which some Fe$^{3+}$ in SrFe$_{12}$O$_{19}$ was replaced by Co$^{2+}$, was evaluated by DFT calculations with HSE06, where $i$ denotes the Fe sites (2a, 2b, 4f$_1$, 4f$_2$ and 12k) occupied by Co.
In all cases, Fe and Co converged to trivalent and divalent high spin states, respectively.
Calculations were carried out for both cases where the moment of substituted Co is in the same direction ($\uparrow \uparrow$) and opposite direction ($\uparrow \downarrow $) to the magnetic moment of Fe in the host (Fe site where Co is substituted).
When Co occupies the majority spin sites (12k, 2a, 2b), $\uparrow \uparrow$ indicates that the Co moment is parallel to the net magnetisation, $\uparrow \downarrow$ indicates that the Co moment is antiparallel to the net magnetisation, and when Co occupies the minority spin sites (4f$_1$, 4f$_2$), note that $\uparrow \uparrow$ indicates that the Co moment is antiparallel to the net magnetisation and $\uparrow \downarrow$ indicates that the Co moment is parallel to the net magnetisation.

The total energy was minimised when Co occupied the 4f$_1$ site with $\uparrow \uparrow$.
The energy difference $\Delta E(i)$ was then defined with respect to $E(\textrm{4f}_1)$ for $\uparrow \uparrow$.
These values are shown in table \ref{table_energy} and figure \ref{fig_Sr_dE}.
First, regardless of the site occupied by Co, the $\uparrow \uparrow$ case has an energy of 100--400 meV lower than the $\uparrow \downarrow$ case.
Therefore, the magnetic moment of Co in the ground state is always stable in the same direction as that of the host Fe.
Although there is no experimental evidence that $A$Fe$_{12}$O$_{19}$ is $\uparrow \uparrow$ in La--Co co-substituted $A$Fe$_{12}$O$_{19}$, the above results provide the first evidence for the assumption of $\uparrow \uparrow$ in this system.
In the following, only the case of $\uparrow \uparrow$ is considered.

The total energy $E(i)$ of Co substituted SrFe$_{12}$O$_{19}$ is $E(\textrm{4f}_1) < E(\textrm{2a}) < E(\textrm{12k}) \ll E(\textrm{4f}_2) < E(\textrm{2b})$.
Thus, Co occupies the 4f$_1$ site in the ground state.
However, as discussed in section \ref{sec_Tdependence}, the cation distribution in a real material freezes at a certain finite temperature.
$\Delta E(\textrm{2b})$ and $\Delta E(\textrm{4f}_2)$ are large of $\sim$1 eV, whereas the synthesis temperature of the M phase never exceeds 2000 K; therefore, Co practically does not occupy 2b or 4f$_2$ sites.
On the other hand, $\Delta E(\textrm{2a})$ and $\Delta E(\textrm{12k})$ are $\sim$100 meV; therefore, if we assume that the cation distribution freezes at approximately 1000 K, Co is distributed in three sites, 4f$_1$, 2a and 12k.
In principle, it is impossible to selectively replace only one specific site with Co that is more or less distributed among the three sites.
The above calculations are in agreement with the experimental results (figure \ref{fig_NMR}), in which Co was distributed over three sites and mainly occupied the 4f$_1$ site.

\begin{figure}[t]
\begin{center}
\includegraphics[width=0.8\textwidth]{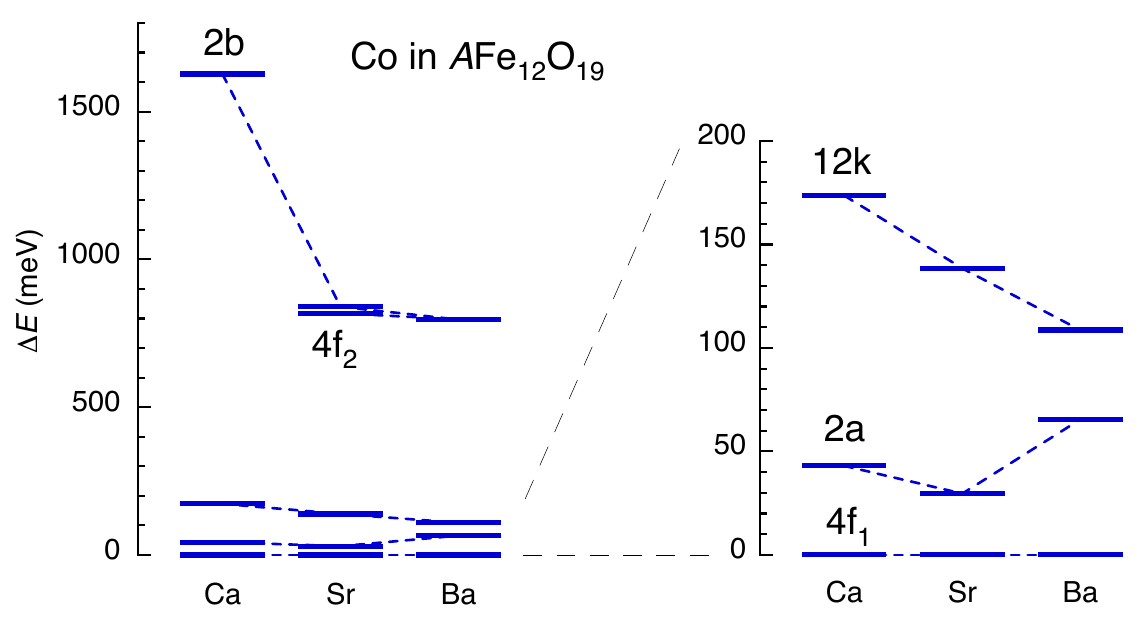}
\caption{\label{fig_dE}
Energy change of $A$Fe$_{12}$O$_{19}$ ($A$ = Ca, Sr and Ba) evaluated by DFT calculations with HSE06 when each Fe site is replaced by Co.
The origin is the energy when Co occupies the 4f$_1$ site.
In the case of $A$ = Ca, a state where Co$^{2+}$ occupies the 4f$_2$ site was not obtained as a stable solution (see text).
The right figure is an enlarged view of the low energy part.
}
\end{center}
\end{figure}

\subsubsection{Co-substituted CaFe$_{12}$O$_{19}$ and BaFe$_{12}$O$_{19}$}
\label{sec_DFT_CaBa}

To test the hypothesis that the Co distribution correlates with the uniaxial strain of the crystal, we performed the same calculations as in the previous section for CaFe$_{12}$O$_{19}$ and BaFe$_{12}$O$_{19}$ with different $A$-ion radii, and compared the results.
We assumed that $\uparrow \uparrow$ is generally more stable than $\uparrow \downarrow$ for La--Co co-substituted $A$Fe$_{12}$O$_{19}$, and for $A$ = Ca and Ba we only calculated in the $\uparrow \uparrow$ case.

Since $E(\textrm{4f}_1)$ was minimal for all $A$ = Ca, Sr and Ba, we defined $\Delta E(i)$ with $E(\textrm{4f}_1$) as the origin, as in the previous section.
The $\Delta E(i)$ values for each system are shown in figure \ref{fig_dE} and table \ref{table_energy}.
In the case of $A$ = Ca, calculations using HSE06 for the supercell with Co substituted for Fe at the 4f$_2$ site show that the Co$^{2+}$ state is unstable. 
In other words, the Co valence converged to a significantly higher state ($\sim$+2.5) even when the Co valence value was initially set to +2, and the Fe valence at the 2a site decreased accordingly. 
As a result, the magnetisation per supercell did not change from the undoped state. 
Therefore the total energy of the 4f$_2$-Co case for $A$ = Ca cannot be compared with the other cases.
The relationship $E(\textrm{4f}_1) < E(\textrm{2a}) < E(\textrm{12k}) \ll E(\textrm{4f}_2) < E(\textrm{2b})$ holds in all cases (except the Ca 4f$_2$-Co case), but the overall trend is that the distribution of $\Delta E(i)$ is largest for $A$ = Ca and smallest for $A$ = Ba.
In all cases, $\Delta E(\textrm{2a})$ and $\Delta E(\textrm{12k})$ are of the order of 100 meV, whereas $\Delta E(\textrm{4f}_2)$ and $\Delta E(\textrm{2b})$ are of the order of 1 eV.
This implies that at finite temperatures of $\sim$1000 K, Co tends to aggregate at the most stable 4f$_1$ site when $A$ = Ca, whereas it tends to be evenly distributed among the three sites when $A$ = Ba.
This trend is in qualitative agreement with the experimental results (figure \ref{fig_NMR}), confirming that the Co distribution is correlated with the uniaxial strain of the crystal.
The quantitative trends are discussed in section \ref{sec_Tdependence}.

\section{Discussion}

\subsection{Orbital contribution to hyperfine fields}
The internal magnetic field at the Co nucleus is given by $H_\mathrm{int} = A_\mathrm{spin} m_\mathrm{spin} + A_\mathrm{orb} m_\mathrm{orb}$, where $A_\mathrm{spin}$ and $A_\mathrm{orb}$ are the spin and orbital components of the hyperfine coupling constants, respectively, and $m_\mathrm{spin}$ and $m_\mathrm{orb}$ are the spin and orbital moments, respectively. 
Since $| A_\mathrm{spin} | \ll | A_\mathrm{orb} |$ in general, when the magnetic moment of Co has an orbital component $m_\mathrm{orb}$,  the resonance frequency depends significantly on the magnitude of $m_\mathrm{orb}$ and cannot be predicted in advance.
Conversely, the frequencies of S1, S2 and S3 differ significantly due to the different magnitudes of $m_\mathrm{orb}$.
The S1 resonance shifts to the high frequency side when an external magnetic field is applied \cite{Sakai2018}.
In other words, an external magnetic field is added to the internal field at the 4f$_1$ site.
Since the 4f$_1$ site is a minority spin site, the magnetic moment of Fe at the 4f$_1$ site is antiparallel to the external field when the sample particles are oriented in the magnetic field.
Furthermore, we assumed that the magnetic moment of the substituted Co is in the same direction as the magnetic moment of the host Fe, considering the results of the DFT calculations (section \ref{sec_Mdirection}).
In this case, the above observation implies that the sign of the internal field at Co occupying the 4f$_1$ site is negative.
However, since the 2a and 12k sites are in octahedral coordination, there may be degeneracy in the ground state orbitals and residual orbital moments.
In general, since $A_\mathrm{orb}$ has a large positive value, the orbital moment produces a relatively large field.
Therefore, the internal fields of S2 and S3 were assumed to be positive, although their sign was experimentally unknown.
The magnitude of the internal field at each site, evaluated on the basis of the above assumptions, is shown in table \ref{table_NMR}.

Assuming $A_\mathrm{spin} = -12$ T/$\mu_\mathrm{B}$ \cite{Watson1961} and $A_\mathrm{orb} = +65$ T/$ \mu_\mathrm{B}$ \cite{Hirosawa1982}, and fixing $m_\mathrm{spin}$ at 3 $\mu_\mathrm{B}$/Co for convenience, the Co orbital moments $m_\mathrm{orb}$ at 4f$_1$ (S1), 12k (S2) and 2a (S3) for the sample with $A$ = Sr are estimated to be approximately 0.4, 1.0 and 1.1 $\mu_\mathrm{B}$/Co.
Although these may be overestimated (if $|A_\mathrm{spin}|$ and $m_\mathrm{spin}$ are overestimated), the relative magnitude appears to be reliable.
These results suggest that relatively large $m_\mathrm{orb}$'s remain in Co occupying the 2a and 12k sites in octahedral coordination, while a small $m_\mathrm{orb}$ is induced in Co at the 4f$_1$ site in tetrahedral coordination through spin-orbit interactions and interactions with ligands.
The orbital field at the 4f$_1$ site is not large enough to cancel the spin component of the internal field, resulting in a relatively small negative internal field.
At the 2a and 12k sites, the orbital field was approximately twice as large as the spin field, resulting in a large positive magnetic field.

According to \cite{Nakamura2019}, only Co at the 4f$_1$ site was effective in enhancing the macroscopic anisotropy.
At first glance, it may seem strange that despite the large orbital moments of Co at the 2a and 12k sites, they do not contribute to the anisotropy enhancement.
However, there is no contradiction if only the Co at the 4f$_1$ site is uniaxially anisotropic and the Co at the 2a and 12k sites is in-plane anisotropic.
Cluster calculations taking into account interactions with ligands show that the sign of the magnetic anisotropy of each site is sensitive to the local strain and can be positive or negative \cite{Inoue2019, Inoue2020}.
Therefore, it is not unnatural for the above situation to be realised in M-type ferrites.
It is concluded that the presence of Co in octahedral coordination is undesirable for enhancing macroscopic uniaxial anisotropy, and that Co at the 2a and 12k sites should be reduced as much as possible to improve performance as a hard magnet.

\begin{figure}[t]
\begin{center}
\includegraphics[width=0.95\textwidth]{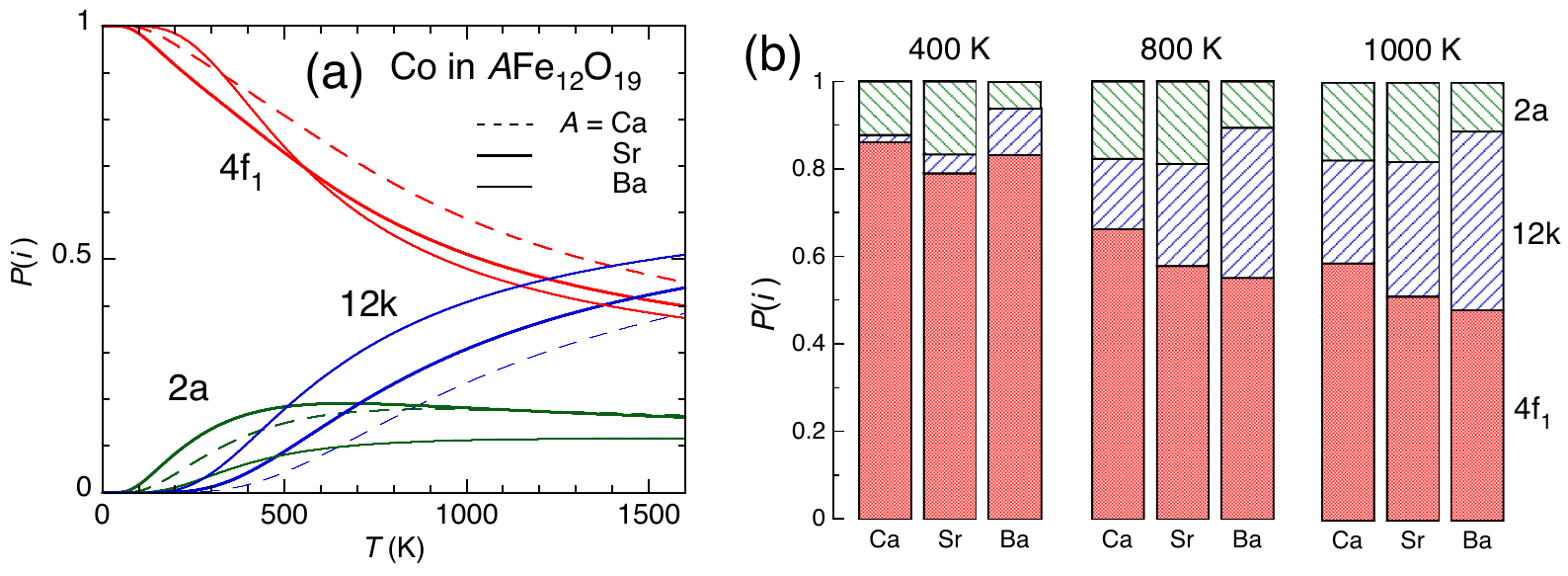}
\caption{\label{fig_P}
(a) Temperature dependence of the probability of Co occupancy at each Fe site evaluated based on the energy difference obtained from DFT calculations with HSE06 for $A$Fe$_{12}$O$_{19}$ ($A$ = Ca, Sr and Ba).
(b) Schematic of Co distribution of $A$Fe$_{12}$O$_{19}$ expected at specific temperatures (400, 800 and 1000 K).
}
\end{center}
\end{figure}

\subsection{Temperature dependence of Co distribution}
\label{sec_Tdependence}

The results of the DFT calculations provided information on the ground state.
In real materials, the cation distribution freezes at a finite temperature.
Therefore, we considered the cation distributions at finite temperatures using the Gibbs free energy.
The Gibbs free energy change when Co occupies site $i$ is given by $\Delta G(i) = \Delta E(i) + p\Delta V(i) - T\Delta S(i)$, where $\Delta E(i)$ is the energy difference in the ground state when Co occupies site $i$, as defined in section \ref{sec_Mdirection}.
$\Delta V(i)$ and $\Delta S(i)$ are the corresponding volume and entropy changes, respectively, $p$ is the pressure, and $k_\mathrm{B}$ is the Boltzmann constant.
The $pV$ term can affect the results, mainly through the thermal expansion of the material.
However, the volume change of SrFe$_{12}$O$_{19}$ between 0 K and 1000 K is approximately 3\% \cite{Fricke2022}, and the change in $pV$ at this time can be roughly estimated to be on the order of 1 neV.
This is sufficiently small compared with the $\Delta E$ we are discussing (on the order of 10--100 meV) to be negligible.
The entropy change $\Delta S(i)$ consists of the contributions of the configuration and the vibrations (phonons, magnons, \textit{etc}).
If the number of sites to be exchanged does not change with temperature, the contribution from the configuration entropy does not affect the statistical average of the physical quantity.
The vibrational contribution is estimated to be approximately 0.1--0.2 $k_\mathrm{B}$/atom for binary substitutional alloys \cite{vandeWalle2002}. 
The corresponding  $T \Delta S(i)$ is of the order of 10 meV at $T = 1000$ K, and the possibility of affecting the total Gibbs energy cannot be ruled out. 
However, the site dependence of $T \Delta S(i)$ due to vibrations is non-trivial and is ignored for simplicity in this analysis.
In this case, the probability $P(i)$ of Co occupying site $i$ at temperature $T$ is determined by $\Delta E(i)$.
Using classical Boltzmann statistics, $P(i)$ is given by
$P(i) = g_i \exp(-\Delta E(i)/k_\mathrm{B} T)/\sum_j g_j \exp(-\Delta E(j)/k_\mathrm{B} T)$, 
where $g_i$ is the multiplicity in the unit cell of site $i$.

Figure \ref{fig_P}(a) shows the temperature dependence of $P(i)$ for each sample.
In the equilibrium state at 0 K, Co should only occupy the 4f$_1$ site.
However, as the freezing temperature of the cation increases, the probability of occupancy of the 2a site in the first excited state $P(\mathrm{2a})$ increases.
Subsequently, the probability of occupancy of the 12k site in the second excited state $P(\mathrm{12k})$ increases.
At relatively high temperatures, $P(\mathrm{2a})$ and $P(\mathrm{12k})$ are reversed .
At temperatures above 1000 K, $P(\mathrm{2a})$ saturates or slowly decreases.
This is because $P(\mathrm{2a})$ approaches the upper limit with a small multiplicity of sites, whereas the 12k sites have a sufficient number of sites.
Close to the upper temperature limit where the M phase is stable, $P(\mathrm{4f}_1)$ and $P(\mathrm{12k})$ are comparable.

As mentioned in section \ref{sec_samplestr}, the samples used in this study may have contained Fe$^{2+}$.
From the charge balance, the amount of Fe$^{2+}$ is approximately 0.04--0.1 relative to 12 Fe.
It is known that Fe$^{2+}$ prefers to occupy the 2a site \cite{Lotgering1974}. 
Therefore, in the equilibrium state at 0 K, Fe$^{2+}$ is expected to occupy at most 10\% of the 2a site.
However, the extra electron associated with Fe$^{2+}$ is expected to hop from site to site as the temperature increases \cite{Chlan2015}. 
The effect is averaged over all Fe sites and should not significantly affect the discussion here.

To observe the $A$-ion dependence in detail, figure \ref{fig_P}(b) shows the Co distribution for each system at specific temperatures (400, 800 and 1000 K).
As a qualitative trend, $P(\textrm{4f}_1)$ is largest for $A$ = Ca and smallest for $A$ = Ba at relatively high temperatures, but this relationship does not necessarily hold at temperatures below 500 K. 
That is, $P(\textrm{4f}_1)$ of BaFe$_{12}$O$_{19}$ increases relatively rapidly at low temperatures.

In actual material synthesis, the cation distribution at a given temperature is not necessarily maintained at room temperature, even if heat treatment is carried out at that temperature.
For example, in the spinel compound MgAl$_2$O$_4$, the time required to equilibrate the ion distribution is a few seconds above 1400 K, but more than 10,000 min below 900 K \cite{Ma2019}.
Assuming that our compounds require a similar duration to equilibrate, at an experimental cooling rate, e.g. $-200$\,$^\circ$C/h, the cation distribution above 1200 K is transformed to a low temperature equilibrium state, whereas at a lower temperature below 900 K the temperature drops before the cation distribution reaches equilibrium.
Therefore, under normal sample preparation conditions, a cation distribution at $\sim$1000 K is maintained at room temperature.
The Co distribution at 1000 K in figure \ref{fig_P}(b) shows that $P$(4f$_1$) is the highest for $A$ = Ca and the lowest for $A$ = Ba.
This trend is qualitatively consistent with the trends observed experimentally (figure \ref{fig_NMR}).
Quantitatively, $P$(4f$_1$) tends to be overestimated in the $^{59}$-Co NMR experiments.
As mentioned in section \ref{sec_NMR}, the experimental error in $P(j)$ seems to be too large to make quantitative comparisons.
Nevertheless, the calculations confirmed that uniaxial compressive strain improves the 4f$_1$ site selectivity of Co.

As mentioned above, $P(\textrm{4f}_1)$ increases as a lower temperature cation distribution is achieved.
Since the time to reach equilibrium is longer at lower temperatures, it is expected that $P(\textrm{4f}_1)$ will increase by post-annealing at lower temperatures or by slowing the cooling rate.
In other words, it is concluded that post-annealing and slow cooling at low temperatures are effective in increasing the uniaxial anisotropy and improving the coercivity and magnetisation of Co-substituted ferrite magnets.
When the temperature variation of $P(\mathrm{4f}_1)$ is steeper in BaFe$_{12}$O$_{19}$ than in SrFe$_{12}$O$_{19}$ and CaFe$_{12}$O$_{19}$, as shown in figure \ref{fig_P}(a), the post-annealing effect is most pronounced in the BaFe$_{12}$O$_{19}$ system.

The site assignment of the S2 and S3 Co-NMR signals is mentioned at the end of this section.
As mentioned above, when a sample is prepared by the usual solid state reaction method, an equilibrium cation distribution of approximately 800--1500 K is frozen, regardless of the cooling rate or post-annealing, taking into account the diffusion rate of the atoms.
Therefore, if we look more closely at the variation of $P(i)$ in the 800--1500 K range in figure \ref{fig_P}(a), the following general trends can be expected.
(i) The temperature variation of $P$(2a) is relatively small and almost independent of the freezing temperature.
(ii) Although the freezing temperature dependence of $P$(4f$_1$) and $P$(12k) is large, the signs of their temperature coefficients are opposite, and the sum of $P$(4f$_1$) and $P$(12k) tends to be constant.
(iii) The $A$-ion dependence is largest at $P$(12k).
These facts suggest that $P$(2a) is relatively insensitive to perturbations such as $A$-ion exchange, freezing temperature variation, and concentration changes, whereas $P$(12k) is expected to be sensitive to these perturbations.
This may be mainly due to the higher multiplicity of sites at the 12k site compared with the 2a site.
Figure \ref{fig_NMR}(b) shows that the experimental $A$-ion dependence was large for S2 and small for S3.
Furthermore, according to the $^{59}$Co-NMR spectrum of La--Co co-substituted SrFe$_{12}$O$_{19}$ reported in \cite{Nakamura2019}, the intensity of S3 hardly changes with the Co concentration, whereas that of S2 decreases significantly with increasing Co concentration (with compression along the $c$-axis).
In the light of the above expected trends and experimental results, it seems reasonable to assign S2 $\rightarrow$ 12k and S3 $\rightarrow$ 2a.
It should be noted that this argument is only an inference based on qualitative trends.
However, this site identification, whether correct or incorrect, does not affect the discussion of the 4f$_1$ site selectivity of Co in this study.

\begin{figure}[t]
\begin{center}
\includegraphics[width=0.8\textwidth]{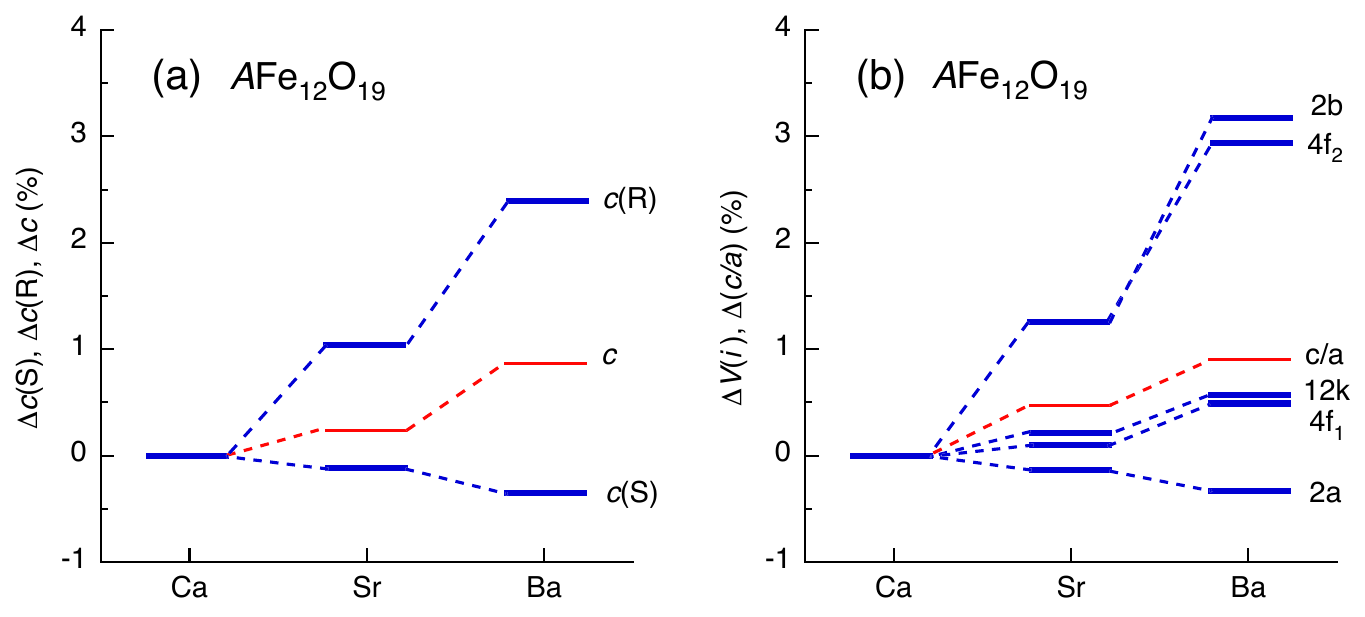}
\caption{\label{fig_localdistortion}
$A$-ion dependence of the relative change in structural parameters of undoped $A$Fe$_{12}$O$_{19}$ with respect to CaFe$_{12}$O$_{19}$ obtained by DFT calculations with HSE06.
(a) R-block length, S-block length and $c$.
(b) Volume of the coordination polyhedron of each Fe site and $c/a$
}
\end{center}
\end{figure}

\subsection{Effect of local structural distortion on Co distribution}
\label{sec_localdistortion}

In section \ref{sec_DFT_CaBa}, we noted that Co distribution correlates with the uniaxial strain of the crystal.
If the energy change associated with the Co substitution (figure \ref{fig_dE}) is determined only by the uniform uniaxial lattice strain, then $\Delta E(i)$ is expected to vary monotonically with the $A$-ion size, and indeed $\Delta E(i)$ decreases monotonically with increasing the $A$-ion size ($c/a$ decreases) at most sites.
However, the 2a site does not follow this trend and $\Delta E(2a)$ varies non-monotonically.
That is, $\Delta E(2a)$ decreases for $A$ = Ca $\rightarrow$ Sr, but conversely increases for $A$ = Sr $\rightarrow$ Ba.
To investigate the cause, we discuss the effect of local deformation in the lattice based on the atomic coordinates of undoped $A$Fe$_{12}$O$_{19}$ evaluated by DFT calculations (table \ref{table_strparam}).

The relative changes in $c$(S) and $c$(R) (with respect to CaFe$_{12}$O$_{19}$) are shown in figure \ref{fig_localdistortion}(a).
As the $A$-ion size increased, $c$(R) increased in response to $c$, but the amount of change was approximately three times that of $c$.
In contrast, $c$(S) showed a negative dependence on the $A$-ion size, but the change was small.
This means that the effect of $A$-ion expansion is not uniform within the crystal and the S block absorbs some of the expansion of the R block.
This may be because the R block to which the $A$ ion belongs is directly affected by the $A$ ion, whereas the S block is not.

Similar non-uniform changes can also be observed in the volume $V(i)$ (table \ref{table_volume}) of the coordination polyhedron at each Fe site.
Figure \ref{fig_localdistortion}(b) shows the relative change $\Delta V(i)$ of $V(i)$ (with respect to CaFe$_{12}$O$_{19}$).
The changes in $c/a$ are included in the same figure for comparison.
The $\Delta V$(2b) and $\Delta V$(4f$_2$) at the R-block sites increased significantly with the expansion of the $A$ ion (increase in $c/a$).
The $\Delta V$(12k) at the R/S block boundary increased only slightly with the expansion of the $A$ ion, which may be due to the combined effect of the large expansion of the R block and the small contraction of the S block, as well as the greater degree of freedom in the position of the oxygen ions forming the 12k coordination polyhedron than at other sites.
In contrast, $\Delta V$(2a) decreased slightly with increasing $A$-ion size.
This is because the height of the 2a coordination polyhedron along the $c$-axis is strongly correlated with the S-block length, as can be seen from the similarity with the relative change in the S-block length $c$(S) in figure \ref{fig_localdistortion}(a).

\begin{table}[t]
\caption{\label{table_Sblock}
Coordinates of oxygen ions forming 2a and 4f$_1$ coordination polyhedra, and aspect ratios of 2a and 4f$_1$ coordinated polyhedra in undoped $A$Fe$_{12}$O$_{19}$ ($A$ = Ca, Sr and Ba) obtained by DFT calculations with HSE06.
}
\begin{indented}
\item[]
\begin{tabular}{ccccccc}
\br
$A$ & \multicolumn{3}{c}{Oxygen coordinate$^a$} & \multicolumn{3}{c}{Aspect ratio$^b$}\\
 & $x$ & $z$ & $z^\prime$ & $r$(2a) & $r$(4f$_1$) & $r$(12k) \\
\mr
Ca & 0.1572 & 0.0527 & 0.0556 & 1.0445 & 0.9848 & 0.9707\\
Sr & 0.1571 & 0.0523 & 0.0553 & 1.0429 & 0.9823 & 0.9677\\
Ba & 0.1568 & 0.0519 & 0.0547 & 1.0407 & 0.9790 & 0.9625\\
\br
\end{tabular}\\
$^a$ Coordinate in O-12k$_1$ $(x, 2x, z)$ and O-4f $(\frac{2}{3}, \frac{1}{3}, z^\prime)$.\\
$^b$ Ratio of the oblique (side length of side triangle) $h$ to the base (side length of base triangle) $b$, $r = h/b$.
$r$(12k) is the average of the different $r$ values in the 12k octahedron.
\end{indented}
\end{table}

\begin{figure}[t]
\begin{center}
\includegraphics[width=0.17\textwidth]{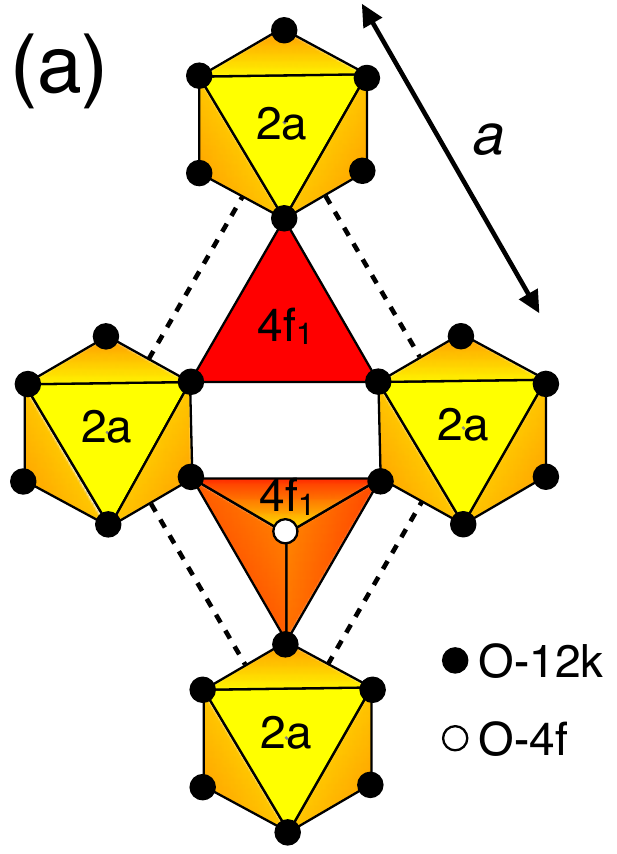}~~~
\includegraphics[width=0.65\textwidth]{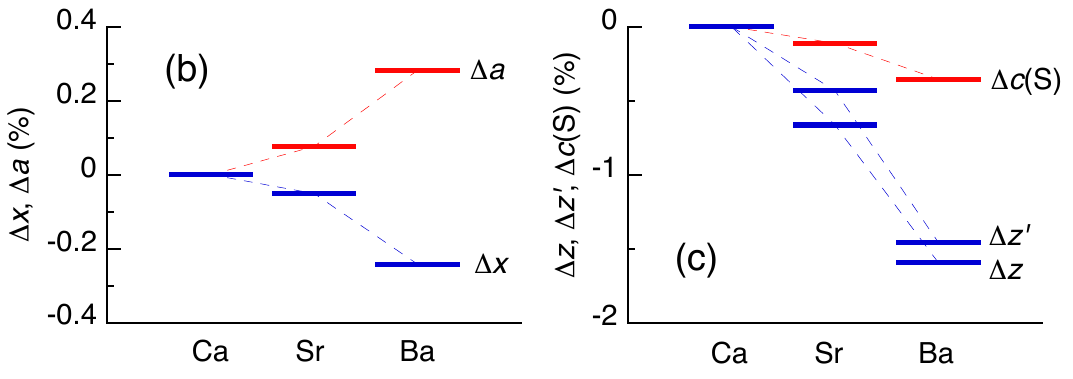}
\caption{\label{fig_Sblock}
(a) 2a and 4f$_1$ coordination polyhedra viewed along the $c$-axis.
Only the oxygen ions are shown.
The atomic coordinates of the ligand oxygen ions (closed circles, O-12k$_1$ site) in the 2a octahedron (yellow) are given by $(x, 2x, z)$.
In this case, the length of the base is $3x \times a$ and the height along the $c$-axis is $2z \times c$.
The vertices of the 2a octahedron are also those of the base triangle of the 4f$_1$ tetrahedron (red). 
The 4f$_1$ tetrahedron has a base length $(1 - 3x) \times a$.
The sum of the base length of the 4f$_1$ tetrahedron and the base length of the 2a octahedron is equal to the lattice constant $a$.
The atomic coordinates of the remaining vertex of the 4f$_1$ tetrahedron (open circle, O-4f site) are given by $(\frac{2}{3}, \frac{1}{3}, z^\prime)$ and have $c$-axis degrees of freedom independent of the 2a octahedron.
The $c$-axis height of the 4f$_1$ tetrahedron is $(z + z^\prime) \times c$.
If the $c$-axis displacements of O-12k$_1$ and O-4f are comparable, the volumes of the 4f$_1$ tetrahedron and the 2a octahedron are determined by the $c$-plane displacement of O-12k$_1$ and show a trade-off tendency if $a$ is invariant.
(b), (c) Relative changes in the coordinates of the oxygen atoms constituting the 2a and 4f$_1$ coordination polyhedra ($x$ and $z$ in O-12k $(x, 2x, z)$ and $z^\prime$ in O-4f $(\frac{2}{3}, \frac{1}{3}, z^\prime))$ and lattice constants ($a$ and S-block length $c$(S)) in undoped $A$Fe$_{12}$O$_{19}$ ($A$ = Ca, Sr and Ba) (CaFe$_{12}$O$_{19}$ basis) obtained by DFT calculations with HSE06.
}
\end{center}
\end{figure}

\begin{figure}[t]
\begin{center}
\includegraphics[width=0.8\textwidth]{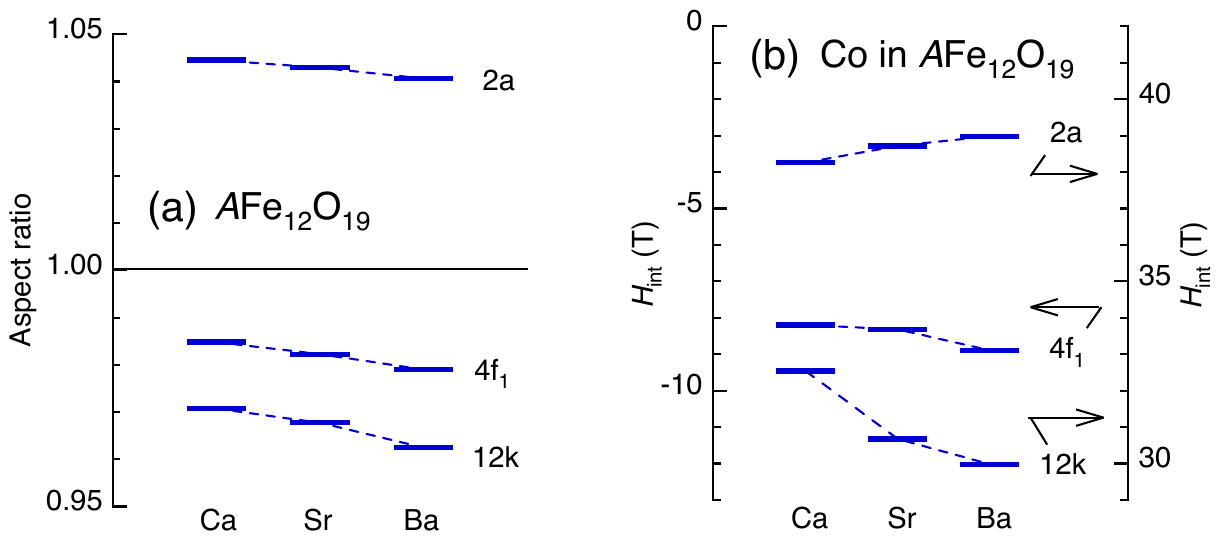}
\caption{\label{fig_AspectRatio}
(a) The $A$-ion dependence of the ratio of the oblique side to the base of the 2a, 4f$_1$ and 12k coordinated polyhedra in $A$Fe$_{12}$O$_{19}$ ($A$ = Ca, Sr and Ba) obtained by DFT calculations with HSE06.
(b) Internal magnetic fields at nuclear sites of Co occupying 2a, 4f$_1$ and 12k sites in La--Co co-substituted $A$Fe$_{12}$O$_{19}$.
}
\end{center}
\end{figure}

It is interesting to note the $A$-ion dependence of the volume of the 4f$_1$ coordination polyhedron belonging to the same S block.
The $A$-ion dependence of $\Delta V$(4f$_1$) (figure \ref{fig_localdistortion}(b)) shows a trend that contradicts the change in the length of the S block (figure \ref{fig_localdistortion}(a)) to which it belongs.
This is due to the fact that $\Delta V(\textrm{2a})$ and $\Delta V(\textrm{4f}_1)$ are not independent.
Figure \ref{fig_Sblock}(a) schematically shows the projection of the S block onto the $c$-plane.
The 4f$_1$ tetrahedron shares the vertices of its base triangle with the 2a octahedron, and the sum of the lengths of the base of the 4f$_1$ tetrahedron and the base of the 2a octahedron is equal to the lattice constant $a$.
The Wyckoff positions of the coordinating oxygens of the 2a octahedron are all 12k$_1$ $(x, 2x, z)$, the length of the base is $3x \times a$ and the height along the $c$-axis is $2z \times c$.
The 4f$_1$ tetrahedron shares the vertices of its base triangle with a 2a octahedron and has a base length of $(1-3x) \times a$.
The remaining vertex of the 4f$_1$ tetrahedron is 4f $(\frac{2}{3}, \frac{1}{3}, z^\prime)$ and has $c$-axis degrees of freedom independent of the 2a octahedron.
In this case, the height of the 4f$_1$ tetrahedron along the $c$-axis is $(z + z^\prime) \times c$.
Thus, if the $c$-axis displacements of the oxygen ions at the 12k$_1$ and 4f positions are comparable, $V$(4f$_1$) and $V$(2a) are determined by the in-plane displacement of the oxygen ion at the 12k$_1$ position.
If $a$ is invariant, $\Delta V$(4f$_1$) and $\Delta V$(2a) show a trade-off tendency.

The coordinates of the oxygen positions 12k$_1$ $(x, 2x, z)$ and 4f $(\frac{2}{3}, \frac{1}{3}, z^\prime)$ in the 2a and 4f$_1$ coordination polyhedra in undoped $A$Fe$_{12}$O$_{19}$ ($A$ = Ca, Sr and Ba) calculated by DFT calculations are given in table \ref{table_Sblock}.
The relative changes with respect to CaFe$_{12}$O$_{19}$ are shown in figures \ref{fig_Sblock}(b) and (c).
The $z$ and $z^\prime$, which correspond to the heights of the coordination polyhedra along the $c$-axis, vary almost similarly to $A$, and the displacement of the oxygen ions along the $c$-axis in the S block is almost uniform.
However, $x$, which corresponds to the length of the sides of the triangle in the $c$-plane, decreases with increasing  $A$-ion size (slightly increasing for $a$).
This means that the base of the 4f$_1$ tetrahedron is enlarged and the base of the 2a octahedron is reduced.
This explains why the $A$-ion dependence of $\Delta V(\textrm{2a})$ and $\Delta V(\textrm{4f}_1)$ show opposite trends, but it is not clear whether the 2a octahedron or the 4f$_1$ tetrahedron is expanding or contracting.

Next we look at the changes in the 2a-octahedron and the 4f$_1$-tetrahedron from a symmetry point of view.
Table \ref{table_Sblock} and figure \ref{fig_AspectRatio}(a) show the ratio $r = h/b$ of the length $b$ of the base (side of a triangle parallel to the $c$-plane) and $h$ of the hypotenuse (side of a side triangle) of each polyhedron, calculated using DFT calculations, where $r = 1$ corresponds to a regular polyhedron.
In the 2a octahedron, $r > 1$, the octahedron is elongated along the $c$-axis, but the stresses due to the $A$-ion expansion cause $r$ to decrease and the 2a octahedron approaches the regular octahedron.
On the other hand, in the 4f$_1$ tetrahedron, $r < 1$, the tetrahedron contracts along the $c$-axis, but the stresses due to $A$-ion expansion reduce $r$ and further increase the distortion.
That is, the 2a octahedron (containing isotropic Fe$^{3+}$) has room to stabilise against $c$-axis stress by increasing symmetry, whereas the 4f$_1$ tetrahedron has no such room.
As a result, the 4f$_1$ tetrahedron expands, and the 2a octahedron, which has a high symmetry and spatial margin, absorbs the distortion, thereby stabilising the whole system.
It can be said that the 2a octahedron plays a buffering role in absorbing strain in the crystal.

Comparing figure \ref{fig_dE} and figure \ref{fig_localdistortion}(b), the change in $\Delta E(i)$ is approximately inversely correlated with the change in $\Delta V(i)$.
In other words, a large positive volume change in the coordination polyhedron tends to increase the probability of Co occupancy.
This may be due to the decrease in elastic energy loss when Co$^{2+}$, which has a larger ionic radius than Fe$^{3+}$, occupies the Fe sites. 
Finally, the non-monotonic $A$-ion dependence of $\Delta E(\textrm{2a})$ shown in figure \ref{fig_dE} can be interpreted as the 2a coordination polyhedron contracting with the expansion of $A$, unlike the other sites.
The fact that there is a trade-off between $V$(4f$_1$) and $V$(2a)  may become apparent during the final optimisation process for the performance of the M-type ferrite magnet.
In conclusion, lattice distortion is not necessarily uniform and differences in local distortion also have a secondary effect on the distribution of Co sites.

We mentioned in section \ref{sec_NMR} that the resonance frequencies of the S1, S2 and S3 resonances (table \ref{table_NMR}) shift slightly depending on the $A$ ion.
The resonance frequencies were converted to internal fields based on the site assignment described above, and are shown in figure \ref{fig_AspectRatio}(b).
The internal field shifted positively at the 2a site and negatively at the 4f$_1$ and 12k sites with increasing $A$-ion size.
The aspect ratio $r$ in figure \ref{fig_AspectRatio}(a) corresponds to the deviation from the cubic symmetry.
The $r$ of the 12k coordination octahedron,  which is the average of the base and oblique sides at the 12k site, has also been added to the figure.
Comparing (a) and (b) in figure \ref{fig_AspectRatio} we can see that as $r$ approaches 1, that is, as the coordination polyhedron approaches cubic symmetry, the internal field shifts in the positive direction.
Assuming that $m_\mathrm{spin}$ is invariant, a positive shift in the internal field corresponds to an increase in $m_\mathrm{orb}$.
Thus, it can be seen that $m_\mathrm{orb}$ tends to increase as the symmetry approaches cubic symmetry.

\section{Summary}

In the La--Co co-substituted M-type ferrite $A$Fe$_{12}$O$_{19}$, Co mainly occupies the 4f$_1$ site (minority spin site) in tetrahedral coordination, and some Co occupies the 2a and 12k sites (majority spin sites) in octahedral coordination.
However, as proposed in \cite{Nakamura2019}, only Co occupying the 4f$_1$ site was found to be effective in enhancing uniaxial anisotropy.
To improve the performance (anisotropy and magnetisation) of M-type ferrite magnets with a limited amount of Co, it is important to concentrate Co at the 4f$_1$ site.
However, their magnetic properties depend on the type and concentration of the $A$ ions, even if the Co content is the same.
Previous studies have shown that the decrease in $c$-axis length due to contraction of the $A$ ion, that is, uniaxial compressive strain, increases the 4f$_1$ site selectivity of Co.
In this study, we performed $^{59}$Co-NMR on La--Co co-substituted M-type ferrites with different $A$ ions, that is, $A$ = Ca, Sr and Ba (ion sizes are Ca$^{2+}$ $<$ Sr$^{2+}$ $<$ Ba$^{2+}$) with Co compositions close to 0.2, and experimentally confirmed the site distribution of Co.
At the same time, DFT calculations were performed for undoped $A$Fe$_{12}$O$_{19}$ and ($A$Fe$_{12}$O$_{19}$)$_8$ with Co$^{2+}$ replacing one Fe atom, [($A$Fe$_{11.875}$Co$_{0.125}$O$_{19}$)$_8$]$^{-}$, and the stable structure and preferential site occupancy of Co were predicted using first-principles calculations.
The results were as follows.

\begin{enumerate}
 
\item 
In La--Co co-substituted $A$Fe$_{12}$O$_{19}$, it was experimentally confirmed that Co concentrates at the 4f$_1$ site as $A$ ions become smaller.
 
\item 
The resonance frequency (resonance field) of Co occupying each Fe site was distributed over a wide frequency (field) range, which was mainly attributed to the difference in the orbital components of the magnetic moment.
The increase in local symmetry at each site correlates with an increase in the orbital field.

\item 
DFT calculations confirmed that when Co occupies the Fe site, its magnetic moment is stable in the same direction as the moment of the host Fe.
 
\item 
According to DFT calculations, the most stable case for any $A$ ion is when Co occupies the 4f$_1$ site, followed by  the 2a and 12k sites with energy differences of the order of 100 meV.
Co at the 2b and 4f$_2$ sites is unstable (the energy difference is of the order of 1 eV) and Co practically does not occupy these sites.
Thus, Co is expected to occupy mainly the 4f$_1$ site, with some Co occupying the 2a and 12k sites, which is in agreement with the experimental results.

\item 
DFT calculations predicted that the energy difference of Co occupying different Fe sites tends to be larger for smaller $A$ ions.
This means that the smaller the $A$ ion, the higher the 4f$_1$ site selectivity for Co, which is in agreement with the experimental results.

\item 
According to the DFT calculations, the strain due to the contraction of the $A$ ions is not uniform throughout the crystal, but has a positive effect (contraction) in the R block and a negative effect (expansion) in the S block.
Therefore, the Co occupancy of each Fe site is not necessarily determined by uniform strain, but is also influenced by differences in the local strain.
In particular, the order of Co occupancy at the metastable 2a and 12k sites may depend on the type of $A$ ion.

\item 
Because the ionic size of Co$^{2+}$ is larger than that of Fe$^{3+}$, the probability of Co$^{2+}$ occupancy generally tends to increase with increasing volume of the oxygen coordination polyhedron at each Fe site.
However, the crystal structure, particularly the S block, is characterised by a trade-off between the volume of the 4f$_1$ coordination polyhedron and that of the 2a coordination polyhedron, which may be important in the final stage of performance optimisation.

\item 
As the cation distribution is frozen at a finite temperature in the actual sample preparation, the 4f$_1$ site occupancy of Co increases as the cation distribution is realised at a lower temperature, taking into account the effect of thermal excitation.
Therefore, post-annealing at low temperature and slow cooling is expected to be effective in improving performance.

 \end{enumerate}

The above results are briefly summarised in terms of guidelines for improving performance (anisotropy and magnetisation) of La--Co co-substituted M-type ferrite magnets with a limited amount of Co. 
It is effective to select $A$ ions with the smallest possible size and to post-anneal at a low temperature or cool slowly to concentrate Co at the 4f$_1$ site in tetrahedral coordination.


\ack
This work was supported by the Elements Strategy Initiative Center for Magnetic Materials (ESICMM), Grant Number JPMXP0112101004, through the Ministry of Education, Culture, Sports, Science and Technology (MEXT), Japan, and JSPS KAKENHI 17K06793, 19K05002 and 22H00263.
The authors would like to thank H. Nishida for his help in preparing Ba-based crystals.

\section*{Conflict of interest}
The authors declare no competing interests.

\section*{ORCID iDs}
Hiroyuki Nakamura https://orcid.org/0000-0001-7085-4800\\
Hiroto Ohta https://orcid.org/0000-0002-0268-9468\\
Takeshi Waki https://orcid.org/0000-0002-3294-9814\\
Yoshikazu Tabata https://orcid.org/0000-0002-4447-2160\\
Hidekazu Ikeno https://orcid.org/0000-0002-3840-4049\\
Christian M\'eny https://orcid.org/0000-0003-0718-161X

\end{document}